\documentclass[printer]{aa}
\usepackage{natbib}
\usepackage{graphicx}
\usepackage[varg]{txfonts}
\usepackage{amsmath} 
\usepackage{bm} 
\usepackage{color} 

\usepackage{colortbl}
\usepackage[]{hyperref}
\usepackage{multirow}
\usepackage{pgf}
\usepackage{tikz}
\usetikzlibrary{decorations.pathmorphing} 
\usetikzlibrary{matrix} 
\usetikzlibrary{arrows,automata} 
\usetikzlibrary{calc} 
\usetikzlibrary{positioning}

\tikzset{
    state/.style={
          rectangle,
           rounded corners,
           draw=black, very thick,
           minimum height=2em,
           inner sep=2pt,
           text centered,
           },
}

\tikzstyle{box}=[rectangle, draw=black, very thick, rounded corners, text width=6cm, inner sep=2pt, minimum height = 2em,
                   rectangle split, rectangle split parts=3]

\newcommand\CP{{\sl Chempy}}

\begin{document}

\title{Chempy: A flexible chemical evolution model for abundance fitting}

\subtitle{Do the Sun's abundances alone constrain chemical evolution models?}

\author{Jan Rybizki\inst{1,2}
\thanks{\emph{Email:} rybizki@mpia.de},
 Andreas Just\inst{2}, Hans-Walter Rix\inst{1}}
\institute{Max Planck Institute for Astronomy, K\"onigstuhl 17, D-69117 Heidelberg, Germany
\and Astronomisches Rechen-Institut, Zentrum f\"ur Astronomie der Universit\"at Heidelberg, M\"onchhofstr. 12-14, D-69120 Heidelberg}

\date{Draft from \today}

\abstract 
{Elemental abundances of stars are the result of the complex enrichment history of their galaxy. Interpretation of observed abundances requires flexible modeling tools to explore and quantify the information about Galactic chemical evolution (GCE) stored in such data. Here we present {\sl Chempy}, a newly developed code for GCE modeling, representing a parametrized open one-zone model within a Bayesian framework. A {\sl Chempy} model is specified by a set of 5-10 parameters that describe the effective galaxy evolution along with the stellar and star-formation physics: e.g. the star-formation history (SFH), the feedback efficiency, the stellar initial mass function (IMF) and the incidence of supernova of type Ia (SN~Ia). Unlike established approaches, {\sl Chempy} can sample the posterior probability distribution in the full model parameter space and test data-model matches for different nucleosynthetic yield sets. It is essentially a chemical evolution fitting tool. We straightforwardly extend \CP~to a multi-zone scheme. As an illustrative application, we show that interesting parameter constraints result from only the ages and elemental abundances of Sun, Arcturus and the present-day interstellar medium (ISM). For the first time, we use such information to infer IMF parameter via GCE modeling, where we properly marginalize over nuisance parameters and account for different yield sets. We find that $11.6^{+2.1}_{-1.6}\,\%$ of the IMF explodes as core-collapse SN (CC-SN), compatible with \citet{Sa55}. We also constrain the incidence of SN\,Ia per $10^3$\,M$_\odot$ to 0.5-1.4. At the same time, this \CP~application shows persistent discrepancies between predicted and observed abundances for some elements, irrespective of the chosen yield set. These cannot be remedied by any variations of {\sl Chempy}'s parameters and could be an indication for missing nucleosynthetic channels. 
\CP\ should be a powerful tool to confront predictions from stellar nucleosynthesis with far more complex abundance data sets and to refine the physical processes governing the chemical evolution of stellar systems.}


\titlerunning{Chempy}
\maketitle

\section{Introduction}
The observed abundances of chemical elements in stars and interstellar medium (ISM) exhibit distinct patterns that correlate with the mass of a galaxy, the position within a galaxy and the birth epoch of the stars \citep{Baade1944,Wallerstein1962}. Such chemical abundance patterns are by far best investigated in the Milky Way (MW), where the detailed photospheric composition of individual stars can be studied through spectroscopy. "Galactic chemical evolution" (GCE) models are a geometrically simplified approach to predicting the temporal evolution of abundance patterns arising from the interplay of the Galaxy's star-formation history (SFH) and star-formation physics with its nucleosynthetic yield, and the inflow and re-processing of the ISM \citep{Tinsley1980,Matteucci2001,Matteucci2012}. The geometric simplification of GCE's in many cases has been the assumption of a "leaky box", or one-zone model, i.e. considering a volume of interest, surrounded by a reservoir and repository of gas.

Since the seminal work of \citet{Schmidt1959}, chemical evolution arguments have been used to infer fundamental Galactic parameters. The one-zone approach was refined to include many elements \citep[e.g.][]{Tinsley1979,Ch97} and was extended to multi-zone models including dynamical constraints \citep{Schonrich2009, Kubryk2015}. This not only increases the parameter space but at the same time makes the models computationally more expensive, in practice prohibiting a full parameter exploration. This places GCE's in between analytical \citep[e.g.][]{Weinberg2016, Spitoni2017} models (with strong simplifying assumption) and hydrodynamical simulations including detailed chemical enrichment as well as galactic dynamics \citep[e.g.][]{Stinson2006, Few2012, Grand2014a}. Because of the complexity of those simulations usually only a limited parameter space or aspect of chemical evolution can be studied \citep[e.g.][]{Jimenez2015}. Simple and flexible GCE on the other hand are ideal to employ large chemical evolution parameter studies.

The current revolution in data quality and quantity on stellar abundances, with the 
far greater ability to constrain stellar ages from spectra \citep{Martig2016,Ness2016,Valls-Gabaud2014,Feuillet2016} calls for a flexible model framework to interpret, and eventually "fit" these data. Recent advances towards this goal have been open-source releases from \citep{Cote2016b,Andrews2017} and statistical measures and sampling techniques to infer e.g. the MW SFH \citep{Snaith2014a} or Sculptor chemical evolution parameters \citep{Cote2017}.

GCE models encompass galactic- (SFH, gas flows), star formation- (IMF, kinetic feedback, SFE) and stellar physics (yields, lifetimes). Modeling, even for only a single-zone, therefore inevitably draws on many free or fit parameters and hyperparameters. Many of them are poorly constrained {\it a priori} \citep{Cote2016b}, yet need to be marginalized out for simple astrophysical inferences. Perhaps the most important external or theoretical input for GCE models are nucleosynthetic yields for the various enrichment channels \citep{Romano2010,Cote2016f}. In the past, theoretical yields have produced mismatches with the observations, leading to the concept of "empirical yields" \citep{Francois2004,Henry2010}. Yet, many abundance trends are not reproduced \citep{Argast2001,Kobayashi2006} and the physical shortcomings of stellar nucleosynthetic yield models are still under debate \citep{Nomoto2013,Fink2014,Pignatari2016,Muller2016}.

In this paper, we lay out an approach to GCE modeling, dubbed {\sl Chempy}, and illustrate its capabilities with a "toy case": trying to match the abundances of the Sun, Arcturus and local B-stars. At its heart, {\sl Chempy} is an open box model, and {\it per se} relatively conventional, the new aspect is the flexible data fitting marginalizing over free parameters and accounting for different yield sets. We also introduce a multi-zone scheme which allows to use several stellar abundances to constrain the same parameters (this will be necessary e.g. when trying to produce empirical yield sets or relax the assumption of a universal IMF).

We begin by introducing the model in Section\,\ref{ch:model}, followed by a data description in Section\,\ref{ch:data}. The Bayesian method will be explained in Section\,\ref{ch:method} and the results including a mock data test and the multi-zone scheme will be presented in Section\,\ref{ch:results}. We will conclude with a summary and an outlook, Section\,\ref{ch:conclusion}.
\section{\CP~and its chemical evolution parameters}
\label{ch:model}
A  Python implementation of the current version of {\sl Chempy}  can be downloaded from \url{https://github.com/jan-rybizki/Chempy}. It was designed to be modular and fast in order to explore a high-dimensional parameter space via Markov Chain Monte Carlo (MCMC).

\begin{figure*}
\begin{tikzpicture}[->,>=stealth']
 \node[align = center] (S) [rectangle,draw=black,very thick,
            rounded corners] 
 {(2) System state $\vec{\mathrm{S}}_{\vec{\theta}\textcolor{gray}{,\vec{\lambda}}}(t)$
};

 \node[align = center, below = 0.5 of S] (P) [rectangle,draw=black,very thick,
            rounded corners] 
 {(3) Predictions$\phantom{_\mathrm{S}}\vec{\mathrm{d}}(\vec{\theta}\textcolor{gray}{,\vec{\lambda}})$
};

 \node[align = center, below = 0.6 of P] (O) [rectangle,draw=black,very thick] 
 {Observations$\phantom{_\mathrm{S}}\vec{\mathcal{O}}$
 
};

\path [<->] ([xshift=1.0cm]P.south) edge node[anchor = east, gray] (L1) [rectangle, draw = gray, very thin, rounded corners, inner sep = 0.08cm,xshift = -0.1cm]{\tiny (4) Likelihood: $\mathrm{P}(\vec{\mathcal{O}}|\vec{\mathrm{d}}(\vec{\theta}\textcolor{gray}{,\vec{\lambda}}))$}  ([xshift=1.0cm]O.north);

 \path (S) edge  (P);

 \node[align = center, above=0.5cm of S] (C) [rectangle,draw=black,very thick,
            rounded corners] 
 {(1) $\mathrm{Chempy(\vec{\theta}\textcolor{gray}{,\vec{\lambda}})}$
};

 \node[align = center, above=0.1cm of C, xshift = -0.815cm, gray] (POSTERIOR) [rectangle, draw = gray, very thin, rounded corners, inner sep = 0.08cm,xshift = -0.1cm]
 {\tiny (5) Posterior: $\mathrm{P}(\vec{\theta}|\vec{\mathcal{O}\textcolor{gray}{,\vec{\lambda}}})$
};

 \node[align = center, right=0.3cm of POSTERIOR, gray] (MCMC) [rectangle, draw = gray, very thin, rounded corners, inner sep = 0.08cm,xshift = -0.0cm]
 {\tiny (6) MCMC sampler
};

 \path (C) edge node[anchor = east, gray, xshift = -0.0cm] (PRIOR) [rectangle, draw = gray, very thin, rounded corners, inner sep = 0.08cm,xshift = -0.1cm] {\tiny (1) Prior: $\mathrm{P}(\vec{\theta})$} (S);

 \path [->] ([xshift = -1.3cm]L1.north) edge[gray] ([xshift = -1.14cm]POSTERIOR.south);
\path [->] ([xshift = -0.60cm]PRIOR.north)  edge[gray] ([xshift = -0.665cm]POSTERIOR.south);
\path [->] (POSTERIOR.east)  edge[gray] (MCMC.west);
\path [->] ([xshift = -0.2cm]MCMC.south)  edge[gray] node[anchor = west, gray, xshift = -0.0cm] {\tiny new $\vec{\theta}$} (C.east);
 \node[right=0.2cm of S,anchor = west] (D0) [rectangle] 
 {:
};

 \node[align = center,right=0.2cm of D0,anchor = west] (S0) [rectangle,draw=black,very thick,
            rounded corners] 
 {$\vec{\mathrm{S}}_{\vec{\theta}}(t=0)$
};

 \node[align = center, below=0.5cm of S0,xshift = 0.4cm] (I) [box,draw=black,very thin,
            rounded corners, gray ,text width=2.3cm,rectangle split parts=2] 
 {\tiny initial conditions
 \nodepart {second} \tiny$\mathrm{m}_\mathrm{SSP}=\mathrm{m}_\mathrm{ISM}=0$\\ \tiny$\mathrm{Z}_\mathrm{corona}=0$\\ \tiny$\mathrm{m}_\mathrm{corona}=$\\
 \tiny$\mathrm{f}_\mathrm{corona}\times\mathrm{m}_\mathrm{SFR}^\mathrm{tot}$
};

 \path (I) edge  ([xshift = 0.4cm]S0.south);

 \node[align = center, right=2.5cm of C,yshift = 0.0cm] (D) [rectangle,draw=black,
            rounded corners, gray] 
 {\tiny stellar \& ISM physics set by parameter $\vec{\theta}$, hyperparameter $\vec{\lambda}$ and Chempy prescriptions
};

 \node[right=0.5cm of S0,anchor = west] (D1) [rectangle] 
 {...
};

 \path (S0) edge  (D1.west);
 
 \node[align = center, right = 0.5cm of D1,anchor = west] (S1) [rectangle,draw=black,very thick,
            rounded corners] 
 {$\vec{\mathrm{S}}_{\vec{\theta}}(\mathrm{t}_\mathrm{p}-\tau_\star)$
};

 \node[align = center, below = 0.5cm of S1] (P1) [rectangle,draw=black,
            rounded corners] 
 {$\left\{\vec{[\mathrm{X}/\mathrm{H}]}_\mathrm{SSP}(\mathrm{t}_\mathrm{p}-\tau_\star)\right\}$
};

 \node[align = center, below = 0.5 of P1] (O1) [rectangle,draw=black] 
 {$\left\{\vec{[\mathrm{X}/\mathrm{H}]}_{\star},\vec{\sigma}_\mathrm{obs}\right\}$
};

\path [<->] (P1) edge  (O1);

 \path (S1) edge  (P1);

 \path (D1) edge  (S1.west);
 
  \node[align = center, right = 0.5cm of S1,anchor = west] (S2) [rectangle,draw=black,very thick,
             rounded corners] 
  {$\vec{\mathrm{S}}_{\vec{\theta}}(\mathrm{t}_\mathrm{p}-\tau_\star+\Delta t)$
 };

  \node[align = center, below = 0.8cm of S2] (A) [rectangle,draw=black, very thin, gray] 
  {\tiny time integration\\ \tiny $\Delta t = 0.1\,\mathrm{Gyr}$\\ \tiny (see Figure\,\ref{fig:mass-flow})
 };
 
 \path (A) edge[gray] ([xshift=0cm]S2.south);
 
 \path (S1) edge (S2);
 
  \node[right=0.5cm of S2,anchor = west] (D2) [rectangle] 
  {...
 };
 
  \path (S2) edge (D2);
  
   \node[align = center, right = 0.5cm of D2,anchor = west] (S3) [rectangle,draw=black,very thick,
              rounded corners] 
   {$\vec{\mathrm{S}}_{\vec{\theta}}(\mathrm{t}_\mathrm{p}=13.5\,\mathrm{Gyr})$
  };

  \node[align = center, below = 0.5cm of S3] (P2) [rectangle,draw=black,
                rounded corners] 
     {$\left\{\vec{[\mathrm{X}/\mathrm{H}]}_\mathrm{ISM}(\mathrm{t}_\mathrm{p}) ,\mathrm{CC}/\mathrm{Ia}(\mathrm{t}_\mathrm{p}),\mathrm{Z}_\mathrm{corona}(\mathrm{t}_\mathrm{p})\right\}$
    };
    
    \path (S3) edge (P2);
    
  \node[align = center, below = 0.5 of P2] (O2) [rectangle ,draw=black] 
  {$\left\{\vec{[\mathrm{X}/\mathrm{H}]}_\mathrm{B-stars},\mathrm{CC}/\mathrm{Ia},\mathrm{Z}_\mathrm{Smith\,cloud},\vec{\sigma}_\mathrm{obs}\right\}$
 };
 
 \path [<->] (P2) edge  (O2);  
  
     \path (D) edge[gray] ([xshift=-0.6cm,yshift=0.1cm]D1);
     \path (D) edge[gray] ([xshift=1.15cm,yshift=0.1cm]S1);
     \path (D) edge[gray] ([xshift=2.5cm,yshift=0.1cm]S2);
    \path (D2) edge (S3);
\end{tikzpicture}
\caption{Schematic summary of {\sl Chempy}: the left portion of this Figure illustrates the sampling of the model parameter posteriors within the Bayesian framework (see Section\,\ref{ch:method}); the right hand portion sketches how {\sl Chempy}  calculates a "system state", for any one set of (hyper-)parameters, which produces observable predictions: (1) for a chosen set of parameters , $\theta$, {\sl Chempy} calculates the system state from initial conditions for all time-steps (2, cf. figure\,\ref{fig:mass-flow}), resulting in the observational predictions (3). These predictions are then compared to a predefined subset of our observations (see Table\,\ref{tab:obs_constraints}); here $\tau_\star$ is the age of the tracers, whose abundance measurements we fit. In our sample application, this is the age of the Sun or Arcturus. We can now calculate the likelihood (4) of any set of observations ($\mathcal{O}$, and their variances $\sigma_\mathrm{obs}$). The posterior (5) is the result of multiplying the likelihood with the parameter priors (see Table\,\ref{tab:abbreviations}). The model parameters' posterior PDF can be sampled using an MCMC algorithm (6). An example of a converged MCMC run can be seen in Figure\,\ref{fig:parameter_space}, where the prior distribution over the parameter space is displayed for comparison.}
\label{fig:time_integration}
\end{figure*}
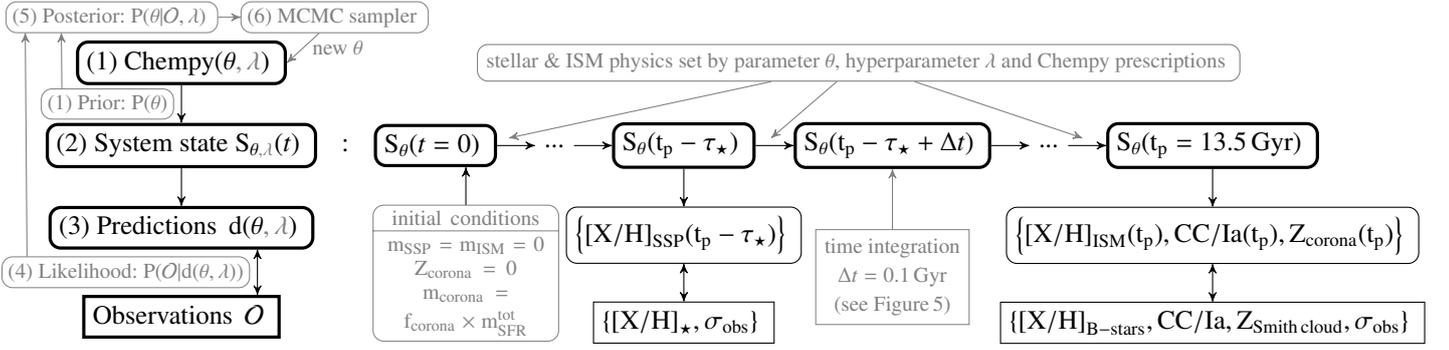

Generally speaking, \CP~is a means of linking the parameters of a chemical evolution model $\vec{\theta}$ (e.g. the IMF high-mass slope) together with underlying hyperparameters $\vec{\lambda}$ (e.g. a specific yield set) to the likelihood of the observations $\vec{\mathcal{O}}$ (e.g. stellar abundances)  via its model predictions $\vec{\mathrm{d}}$. This is schematically shown in Figure\,\ref{fig:time_integration}, which can provide guidance through the methodological description.

We start by introducing the main parameters, ($\vec{\theta}$), needed to specify the chemical evolution model (see Table\,\ref{tab:abbreviations}). Each of these is parameterizing a specific aspect of the GCE model. This list of parameters is of course not exhaustive (e.g. it could include 
the mass of the most massive stars, $\mathrm{m}_\mathrm{max}$), but we limit ourselves to $\mathrm{n}_{\vec{\theta}}=7$ parameters, for the sake of simplicity. We distinguish between parameters governing the physics of the stellar component ($\mathrm{n}_{\vec{\theta}_\mathrm{SSP}}=3$, global parameters) and parameters that affect the ISM ($\mathrm{n}_{\vec{\theta}_\mathrm{ISM}}=4$, local parameters).

Each parameter (in some cases its logarithm) has a Gaussian prior distribution assigned to it, based on broad insights from published work (Table\,\ref{tab:abbreviations}). The prior is specified by the mean $\overline{\vec{\theta}}_\mathrm{prior}$ (i.e. the maximum or peak of the prior distribution) and the standard deviation $\vec{\sigma}_\mathrm{prior}$. In the following we will illustrate how each parameter affects {\sl Chempy} by plotting its functional form or resulting predictions for a range of parameter values.

\begin{tiny}
\begin{table*}
\begin{minipage}{\textwidth}
\begin{center}
\caption{Free \CP~parameter, $\vec{\theta}$, and their priors with assumed Gaussian error model.}
\begin{tiny}
\begin{tabular}{ rl|ccc }
$\vec{\theta}$ & description & $\overline{\theta}_\mathrm{prior}\pm\sigma_\mathrm{prior}$ & limits & approximated prior based on \\
  \hline
  \multicolumn{5}{c}{stellar (SSP) evolution parameters}\\
\hline
  $\alpha_\mathrm{IMF}$ & high-mass slope of the \citet{Chabrier2001} IMF (eq.\,\ref{eq:imf}) & $-2.29\pm0.2$ & $[-4,-1]$ & \citet[tab.\,7]{Cote2016b} \\
  $\log_{10}\left(\mathrm{N}_\mathrm{Ia}\right)$ & number of SN\,Ia exploding per M$_\odot$ over 15\,Gyr & $-2.75\pm0.3$ & $[-\infty,0]$ & \citet[tab.\,1]{Maoz2012a}\\
   $\log_{10}\left(\tau_\mathrm{Ia}\right)$ & SN\,Ia delay time in Gyr for \citet{Maoz2010} distribution & $-0.8\pm0.3$ & $\left[-\infty,1\right]$ & estimate from \cite{Maoz2012} \\
\hline
  \multicolumn{5}{c}{ISM evolution parameters}\\
\hline
  $\log_{10}\left(\mathrm{SFE}\right)$ & star formation efficiency governing the infall and ISM gas mass & $-0.3\pm0.3$ & $[-\infty,\infty]$ & \cite{Bigiel2008} \footnote[1]{{\tiny Theoretical work by \citet{Cote2017} derives values in a range of 2 - 0.03 per Gyr. The work of \citet{Chiappini2001} and \citet{Andrews2017} use 1 per Gyr. \citet{Cote2017} and \citet{Andrews2017} both assume a linear Schmidt law $\mathrm{n}_\mathrm{Schmidt}=1$ (same as this work), whereas \citet{Chiappini2001} uses $\mathrm{n}_\mathrm{Schmidt}=1.5$ together with a gas density threshold. See \citet{Vincenzo2016b} for a detailed comparison between linear and non-linear Schmidt law.}} \\
  SFR$_\mathrm{peak}$ & peak of SFR in Gyr (scale of $\gamma$-distribution with k=2, eq.\,\ref{eq:SFR})& $\phantom{-}3.5\pm1.5$ & $[0,\infty]$ & inspired by \citet[fig\,4b]{VanDokkum2013b} \\
  x$_\mathrm{out}$ & fraction of stellar feedback outflowing to the corona & $\phantom{-}0.5\pm0.2$ & $[0,1]$ & estimate because uncommon parametrization \\
    $\log_{10}\left(\mathrm{f}_\mathrm{corona}\right)$ & corona mass factor times total SFR gives initial corona mass & $\phantom{-}0.3\pm0.3$ & $[-\infty,\infty]$ & \cite{Stern2016}, \cite{Werk2014}
  \label{tab:abbreviations}

\end{tabular}
\end{tiny}
\end{center}
\end{minipage}
\end{table*}
\end{tiny}
\subsection{Stellar physics parameters}
{\sl Chempy}'s central module calculates the yield for a simple stellar population (SSP). This routine is governed by three parameters, which set the stellar physics of the chemical evolution model.

\label{ch:SSP}
The stellar component of  {\sl Chempy} is modeled as a composite stellar population (CSP), a sum of simple stellar populations (SSPs), separated equidistantly in time. 
An SSP is fully characterized by its time of birth, its mass and element composition $\mathrm{SSP}\left(\mathrm{t}_\mathrm{birth},\mathrm{mass},\vec{[\mathrm{X/H}]}\right)$, 
which also fixes its feedback when assuming some IMF, stellar lifetimes and nucleosynthetic yields. The total mass of an SSP for a specific time-step is determined by the star formation rate (SFR) and its initial elemental abundance is given by the composition of the ISM at that time. 
 
\begin{figure}
\resizebox{\hsize}{!}{\includegraphics{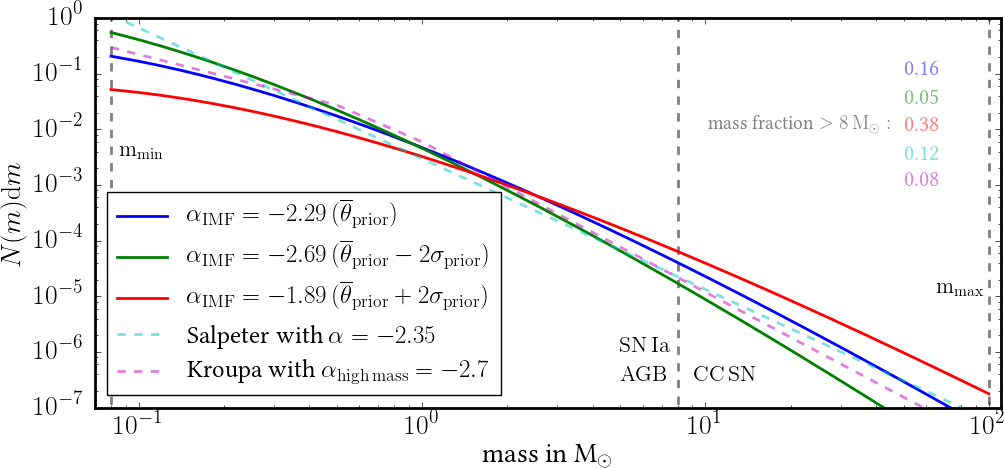}}
\caption{($\vec{\theta}_1=\alpha_\mathrm{IMF}$), showing the high-mass slope of the IMF. Illustration of the number of stars per mass interval. We use a \citet[tab.1, IMF\,3]{Chabrier2001} functional form as in Equation\,\ref{eq:imf} with $\alpha_\mathrm{IMF}=-2.29$ as $\overline{\theta}_\mathrm{prior}$. $\pm2\sigma_\mathrm{prior}$ from $\overline{\theta}_\mathrm{prior}$ are shown (see Table\,\ref{tab:abbreviations}). For comparison the \citet{Kroupa1993} and \cite{Sa55} IMF are depicted. The mass fraction of each IMF that explodes as core-collapse supernova (CC-SN) is written in the top right.}
\label{fig:IMF}
\end{figure}
The stellar masses are distributed on a constant grid from 0.08 to 100\,M$_\odot$ with a mass step that can be adjusted. We found a mass resolution of about $0.02$\,M$_\odot$ to be sufficient. The functional form of the IMF is \citet[tab.1, IMF\,3]{Chabrier2001}
\begin{equation}
\label{eq:imf}
\frac{\mathrm{d}n}{\mathrm{d}m}=m^{-\left(1+\alpha_\mathrm{IMF}\right)}\exp\left[\left(-\frac{716.4}{m}\right)^{0.25}\right] 
\end{equation}
 where the high-mass slope ($\alpha_\mathrm{IMF}=\vec{\theta}_1$) is one of \CP's basic parameters; we assume the IMF slope to be universal. The parameter $\alpha_\mathrm{IMF}$ is crucial, as it sets the ratio of low-mass to high-mass stars and it also alters the number distribution of the high-mass stars which influences the elemental composition of the feedback. The range of IMFs, spanned by this parametrization, is shown in Figure\,\ref{fig:IMF} where we plot it for the mean prior value ($\overline{\theta}_{\mathrm{prior},1}=\alpha_\mathrm{IMF}=-2.29$) and its 2$\sigma_\mathrm{prior}$ deviations, also comparing to the \cite{Sa55} and the \citet{Kroupa1993} IMF. The values of the prior are taken from \citet[tab.\,7]{Cote2016b}, albeit their high-mass slope parameter is not exactly applicable to our IMF functional form. The stellar lifetimes are calculated according to \cite{Argast2000} in order to have the mass and the mass-range of dying stars for all remaining time-steps of the simulation.
 
 We differentiate between three nucleosynthetic channels: Supernova of type Ia (SN\,Ia), core-collapse supernova (CC-SNe) and asymptotic giant branch stars (AGB). For the latter two the elemental feedback and remnant mass depend on the mass of dying stars (see Figure\,\ref{fig:yield}) and we assume that all relevant feedback materials of a star (including winds) is ejected only at the end of its lifetime. The elemental feedback is calculated according to yield tables from literature. For our default yield set (see Table\,\ref{tab:yield}) the AGB feedback is calculated according to \cite{Karakas2010}; for the CC-SN feedback we use the table and prescription of \cite{Nomoto2013} where 50\,\% of CC-SN more massive than 25\,M$_\odot$ explode as Hypernova. We use the net-yields (i.e. only the newly synthesized material appears in the table and the missing ejecta mass is filled with unprocessed material from the stellar birth elemental composition, which is the predicted model ISM composition at the formation time of the corresponding SSP) which are calculated for a grid of masses and metallicities. The interpolation scheme can be switched from linear to logarithmic in metallicity and we use the latter here.

\begin{table*}
\begin{minipage}{\textwidth}

\begin{center}
\caption{Yield sets for which we test our inference}
\begin{tabular}{ l|c cc}
Yield set & CC-SN & SN\,Ia & AGB \\
\hline
Default &\citet{Nomoto2013} (net)
\footnote{\tiny'Net' means that the original yield tables provide only the newly produced material; any material that was originally present in the star and expelled into the ISM without further processing is computed by us according to the chemical composition of the ISM at star's birth predicted by our GCE model.}
 &  \citet{Seitenzahl2013} & \citet{Karakas2010} (net)$^{\,a}$\\
Alternative & \citet{Chieffi2004} (gross)
\footnote{\tiny'Gross' means that we are using the total (newly produced + unprocessed) stellar ejecta provided in the original yield tables.}
 &  \citet{Thielemann2003}& \citet{Ventura2013} (net)$^{\,a}$

  \label{tab:yield}
\end{tabular}
\end{center}
\end{minipage}

\end{table*}

\begin{figure}
\resizebox{\hsize}{!}{\includegraphics{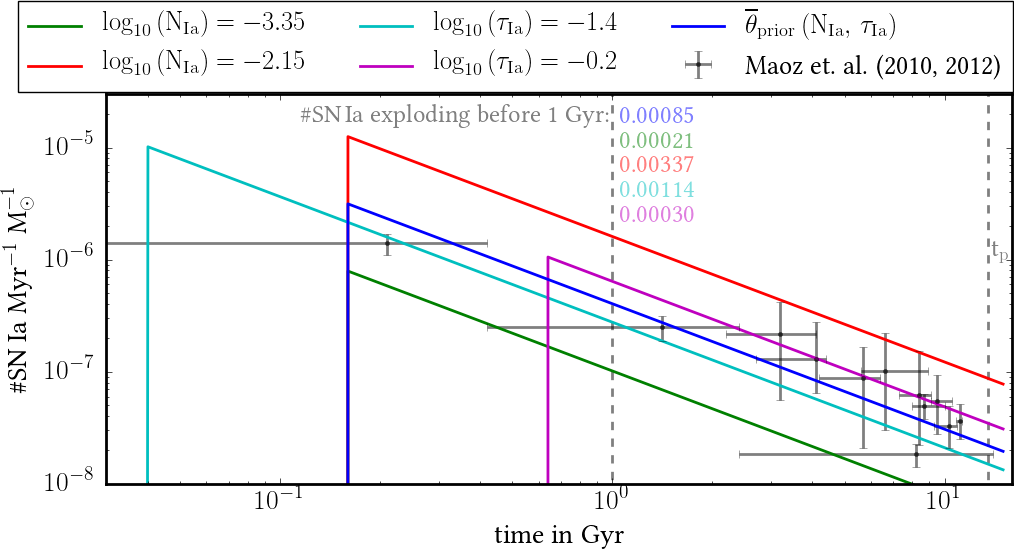}}
\caption{Impact of the model parameters $\vec{\theta}_{2,3}=\mathrm{N}_\mathrm{Ia},\tau_\mathrm{Ia}$, the SN\,Ia normalisation and the SN\,Ia delay time. We show the distribution functions of SN Ia explosions for an SSP of 1\,M$_\odot$, using a \citet{Maoz2010} functional form with a mass normalisation of $\mathrm{N}_\mathrm{Ia}=0.00178$ and a time-delay of $\tau_\mathrm{Ia}
\approx160\,$Myr. We show $\pm2\sigma_\mathrm{prior}$ variations from $\overline{\theta}_\mathrm{prior}$ (see Table\,\ref{tab:abbreviations}). For comparison the \citet{Maoz2010} and \citet{Maoz2012} observations are depicted. The number of SN Ia exploding before 1\,Gyr is indicated as well.}
\label{fig:DTD}
\end{figure}

For the SN~Ia we use the \citet{Seitenzahl2013} yields (their model N100, which best reproduces observables \citep{Sim2013}, without metallicity dependence, and in \CP~SN~Ia always explode with the same mass), calculated from 3D models superseding the W7 model of \cite{Iwamoto1999}, which was calculated in 1D and had old electron capture rates. Because of that {\sl Ni} was over- and {\sl Mn} underproduced, which is remedied with \cite{Seitenzahl2013}. The choice of a specific yield set can be treated as a hyperparameter and we will test the impact that using different yield sets has on our inference.

\begin{figure*}
\resizebox{\hsize}{!}{\includegraphics{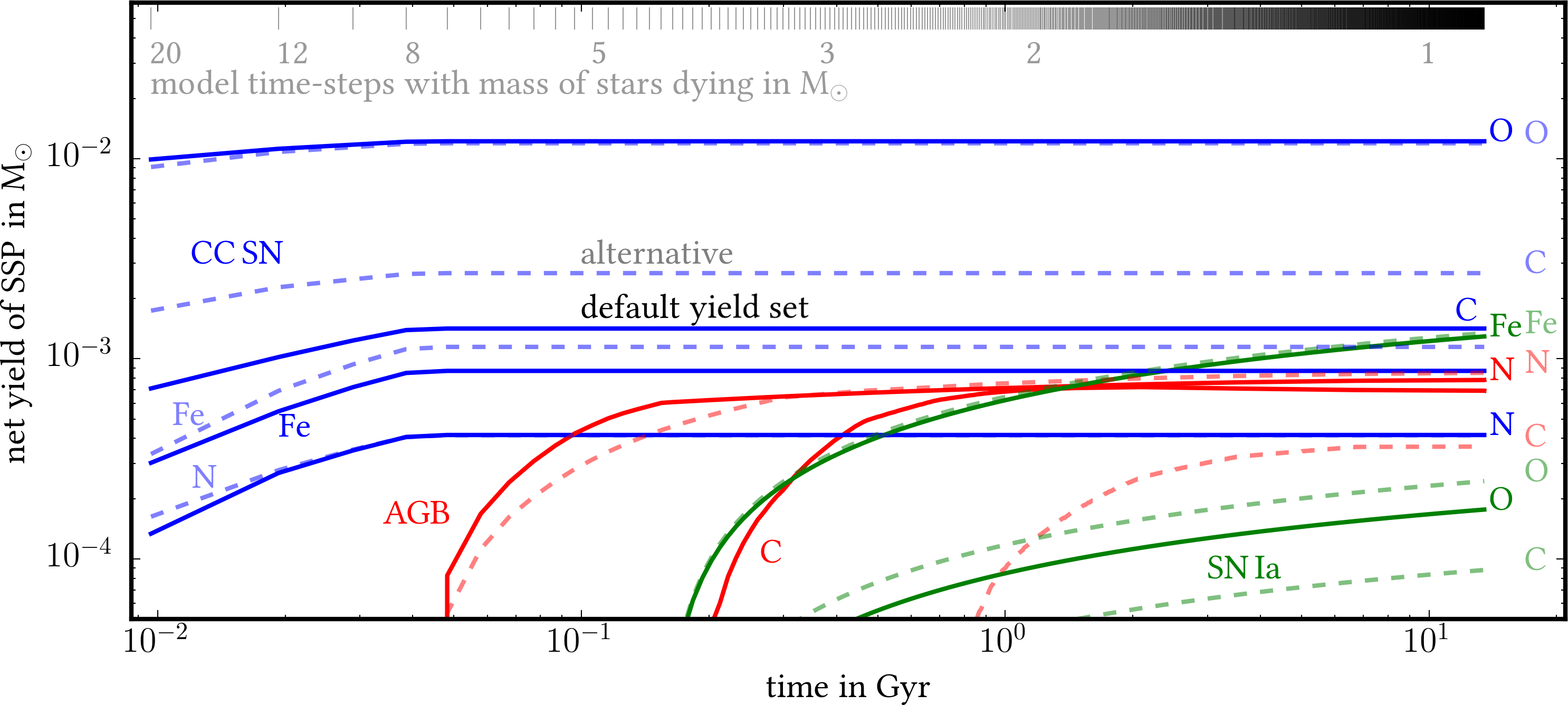}}
\caption{Cumulative net yield in M$_\odot$ for C, N, O and Fe over time for an SSP of 1\,M$_\odot$ and Solar metallicity Z$_\odot$, using Chempy's $\overline{\theta}_\mathrm{prior}$ parameters (see Table\,\ref{tab:abbreviations}). The nucleosynthetic enrichment of CC-SN in blue, SN\,Ia in green and AGB in red are plotted. We show the default yield set in solid lines and for the alternative yieldset in dashed lines (those are defined in Table\,\ref{tab:yield}). Linear time-steps of 9.6\,Myr (in contrast to the fiducial $\Delta\mathrm{t}=0.1$\,Gyr) are indicated by the short vertical lines at the top together with the mass of stars dying at that time. Note, that for the alternative CC-SN yield only gross yields are available (unprocessed material is included in the feedback, see text).}
\label{fig:yield}
\end{figure*}
Which fraction of stars and in what mass range explode as SN\,Ia, is less well understood, and we treat this empirically. The delay time distribution (DTD) of the SN\,Ia explosions are parametrized as in \citet{Maoz2010} with a power-law of t$^{-1.12}$. Free parameters are the number of SN\,Ia per Solar mass over a time of 15\,Gyr ($\mathrm{N}_\mathrm{Ia}=\vec{\theta}_2$) and the time delay for the first SN\,Ia events to occur ($\tau_\mathrm{Ia}=\vec{\theta}_3$). In Figure\,\ref{fig:DTD} the 2$\sigma$ variation of those parameters are shown compared to the default model and the data on SN\,Ia explosions by \citet[tab.\,1]{Maoz2010} and \citet[tab.\,2]{Maoz2012}. The prior of the SN\,Ia normalisation ($\mathrm{N}_\mathrm{Ia}$) is based on \citet[tab.\,1]{Maoz2012a} and the prior on the time delay ($\tau_\mathrm{Ia}$) is estimated from the bin size of \citet{Maoz2012} data.

The distribution of stars along the IMF and the SN~Ia explosions can be calculated as stochastic processes in \CP . But this converges rapidly to the analytic solution, as soon as the SSP masses reach $\sim$$10^5$ M$_\odot$, therefore we employed the faster analytic version here. 

The yield is illustrated in Figure\,\ref{fig:yield} for an SSP of 1\,M$_\odot$ and Solar metallicity 
with $\overline{\vec{\theta}}_\mathrm{prior}$ parameters comparing the default yield set with the alternative yield set (i.e. \citet{Chieffi2004}, \cite{Ventura2013} and \cite{Thielemann2003} for CC-SN, AGB and SN\,Ia respectively), where the cumulative yield over time for O, Fe, N and C are displayed for the three different nucleosynthetic feedback processes\footnote{Additional yield tables implemented in \CP~are: \citet{Portinari1998} (gross), \citet{Pignatari2016} (gross) for CC-SNe, \citet{Iwamoto1999} for SN\,Ia and \citet{Pignatari2016} (gross, only provide Solar and half Solar metallicity, which is not enough for our simulation), \citet{Karakas2010} (gross), \citet{Karakas2016} (net) for AGB stars.}.

This feedback table is stored for each SSP of the simulation on the grid of the remaining time-steps such that the feedback material of the previous stellar generations can be added to the respective latest time-step by simple matrix calculation. Together with the effective use of {\sl numpy} arrays \citep{VanderWalt2011} this diminishes the time per \CP~evaluation to the order of seconds. 
\subsection{ISM physics - mass flow and time evolution}
The mass flow of the open box model is sketched out in Figure\,\ref{fig:mass-flow}. The open box consists of a well-mixed gas-phase representing the ISM from which new stars are formed. So far \CP~ does not separate the warm and atomic gas phase, even though our linear Schmidt law only holds for the molecular gas component \citep{Bigiel2008}. The ISM infall is fed from a well-mixed "corona" gas reservoir, which is slowly enriched by the stellar feedback and on the other hand diluted by inflow of primordial gas. Each time step consists of the following intermediate time steps:
\begin{figure*}
\begin{tikzpicture}[->,>=stealth']
 \node[align = center] (STARS) [box,text width=9.0cm] 
 {\textbf{Stellar component (CSP)}
  \nodepart[align = left,gray] {second} \tiny composite stellar population CSP = multiple simple stellar populations SSP\\\tiny i.e.  $\mathrm{CSP}(t)=\sum_{\mathrm{t}=0}^{\mathrm{t}=t}\mathrm{SSP}\left(\mathrm{t}),\quad\mathrm{SSP}(\mathrm{t}_\mathrm{birth},\,\mathrm{m},\,\vec{[\mathrm{X/H}]}\right)$, derived quantities:\\
  \tiny IMF ($\alpha_\mathrm{IMF}$) \& stellar lifetimes $\Rightarrow\mathrm{m}_\mathrm{dying}(t+\Delta t)$\\
  \tiny together with SN\,Ia DTD ($\mathrm{N}_\mathrm{Ia},\tau_\mathrm{Ia}$) $\Rightarrow$ event rates: $\#_\mathrm{CC,Ia,AGB}(t+\Delta t)$\\
  \tiny yield tables $\Rightarrow\mathrm{m}_\mathrm{feedback_\mathrm{CC,Ia,AGB}}(t+\Delta t)+\mathrm{m}_\mathrm{remnants_\mathrm{CC,Ia,AGB}}(t+\Delta t)$\\ \normalsize
  $\textcolor{black}{\mathrm{CSP}(t+\Delta t)=\mathrm{CSP}(t)+...}$
  
   \nodepart[align = left] {third} \normalsize (2): $-\,\mathrm{m}_\mathrm{feedback}(t+\Delta t)$\\
   (4): $+\,\mathrm{SSP}(t+\Delta t)$
};

 \node[state,    	
  text width=7cm, 	
  yshift=-2.2cm, 		
  right of=STARS, 	
  node distance=8.5cm, 	
  anchor=center] (ISM) [box] 	
  {\textbf{ISM gas (well mixed)}
  \nodepart[align = left] {second} $\mathrm{m}_\mathrm{ISM}(t+\Delta t)$ = $\mathrm{m}_\mathrm{ISM}(t)+...$
   \nodepart[align = left] {third} (2): $+\,(1-\mathrm{x}_\mathrm{out})\times\mathrm{m}_\mathrm{feedback}(t+\Delta t)$\\
   (3): $+\,\mathrm{m}_\mathrm{infall}(t+\Delta t)$\\
      $\phantom{\mathrm{(3):\,}}$\textcolor{gray}{\tiny this fixes $\vec{[\mathrm{X}/\mathrm{H}]}_\mathrm{ISM}(t+\Delta t)$}\\
      
   (4): $-\,\mathrm{m}_\mathrm{SFR}(t+\Delta t)$
 };
 \node
 [state,
  above of=ISM,
  node distance=4.5cm,
  anchor=center,
  text width=7cm] (CORONA) [box]
 {%
  \textbf{Corona gas (well mixed)}
  \nodepart[align = left] {second} $\mathrm{m}_\mathrm{corona}(t+\Delta t)$ = $\mathrm{m}_\mathrm{corona}(t)+...$
   \nodepart[align = left] {third} (1): $+\,\mathrm{m}_\mathrm{inflow}(t+\Delta t)$\\
   (2): $+\,\mathrm{x}_\mathrm{out}\times\mathrm{m}_\mathrm{feedback}(t+\Delta t)$\\
   $\phantom{\mathrm{(2):\,}}$\textcolor{gray}{\tiny this fixes $\vec{[\mathrm{X}/\mathrm{H}]}_\mathrm{corona}(t+\Delta t)$}\\
   (3): $-\,\mathrm{m}_\mathrm{infall}(t+\Delta t)$
 };
\node[state, left of=CORONA, yshift=0.5cm,node distance=7.0cm] (INFLOW)
{
  \textbf{Primordial gas}
};
\node[state, left of=INFLOW, yshift=0cm,node distance=4.0cm,anchor=center, text width = 3cm] (LEGEND) [box,text width = 4.0cm,gray]
{
  \textbf{Chempy - mass flow}
    \nodepart[align = left] {second} System state $\vec{\mathrm{S}}(t)$
     \nodepart[align = left] {third} Time evolution terms $\vec{\mathrm{S}}(\Delta t)$
};
	 \node[ below = 0.0cm, right = 0.5cm] at (STARS.east) {(2) Feedback};
 \path ([yshift = -0.05cm]STARS.east)     	edge[bend left=0] node[xshift = 0.8cm,yshift=0.1cm, gray]{$(1-\mathrm{x}_\mathrm{out})$} ([xshift=-2.3cm]ISM.north);

  \path ([yshift=0cm]ISM.west) 	edge[bend left=0]  
  node[below, xshift = 0cm, yshift = 0.cm]{(4) SFR} 
  node[below, xshift = -0.5cm, yshift =-0.5cm,gray] {$\mathrm{SSP}(t+\Delta t,\mathrm{m}_\mathrm{SFR}(t+\Delta t),\vec{[\mathrm{X}/\mathrm{H}]}_\mathrm{ISM}(t+\Delta t))$}
  ([xshift=0cm]STARS.south);
 \path ([yshift=0.0cm]STARS.east) 	edge[bend right=0]  node[xshift = 0.8cm,yshift=-0.1cm, gray]{$\mathrm{x}_\mathrm{out}$} ([xshift=-2.3cm]CORONA.south);
 \path ([xshift=0.0cm]CORONA.south) edge 
 node[right, yshift=0.2cm,xshift=0.5cm]{(3) Infall} 
 node[right, yshift=-0.5cm,gray]{$\frac{\mathrm{m}_\mathrm{SFR}(t+\Delta t)}{\mathrm{SFE}}-\mathrm{m}_\mathrm{ISM}(t)$} ([xshift=0.0cm]ISM.north);
 \path (INFLOW.east) 	edge[]  
 node[above]{(1) Inflow} 
 node[below,gray]{$1\times\mathrm{m}_\mathrm{SFR}(t+\Delta t)$}
 ([yshift = 0.5cm]CORONA.west);
\end{tikzpicture}
\caption{\CP~(one-zone, open-box) mass flow of one time-step $t\rightarrow t+\Delta t$, illustrating how \CP~is integrating over time from the initial system state, S$_\theta(t=0)$ (see Figure\,\ref{fig:time_integration}). Each box represents a subsystem with its current state and the changes to the next time-step. The numbered arrows show the sequence of \CP~time-integration and the mathematical prescriptions with their parameter dependence. The quantities characterizing the resulting CSP (see Section\,\ref{ch:SSP}) are given in gray. The chemical composition [X/H] of each subsystem is tracked. Initial conditions are: no stars, no ISM gas and $\mathrm{f}_\mathrm{corona}\times\mathrm{m}_\mathrm{SFR}^\mathrm{tot}$ of primordial gas in the corona. In each time-step the inflow of primordial gas into the corona is calculated first (1). Then the feedback from all preceding SSPs is distributed among corona and ISM (2). Next \CP~incorporates enough gas from the corona into the ISM to satisfy $\mathrm{SFR}=\mathrm{SFE}\times\mathrm{m}_\mathrm{ISM}$ (3). This results in a new SSP forming at that time-step from the ISM (4).}
\label{fig:mass-flow}
\end{figure*}
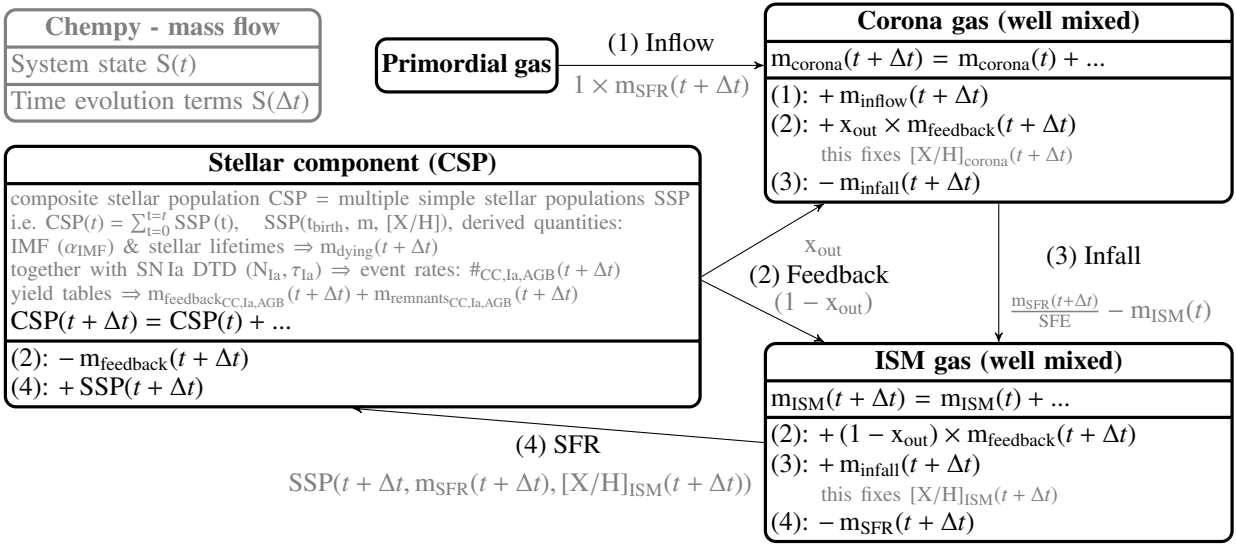

\begin{enumerate}
\item Inflow of primordial gas into the corona. The inflow mass per time-step is, somewhat {\it ad hoc}, set to equal the SFR, $\mathrm{m}_\mathrm{inflow}(t)=\mathrm{m}_\mathrm{SFR}(t)$.
\item Stellar feedback material from previous stellar generations is added up for the particular time-step and the "outflow fraction" ($\mathrm{x}_\mathrm{out}=\vec{\theta}_6$) is presumed to add to the corona. The remaining fraction is mixed into the local ISM. This determines the new corona abundances.
\item Infall from the corona until enough gas is in the ISM so that  $\mathrm{m}_\mathrm{SFR}=\mathrm{SFE}\times\mathrm{m}_\mathrm{ISM}$,
given the star formation efficiency (SFE) parameter ($=\vec{\theta}_4$). This sets the ISM abundances.
\item New stars form with abundances that equal those of the ISM, and their individual masses are distributed following the IMF; the total amount of new stellar mass is set by the prescribed  SFR.
\end{enumerate}
\begin{figure}
\resizebox{\hsize}{!}{\includegraphics{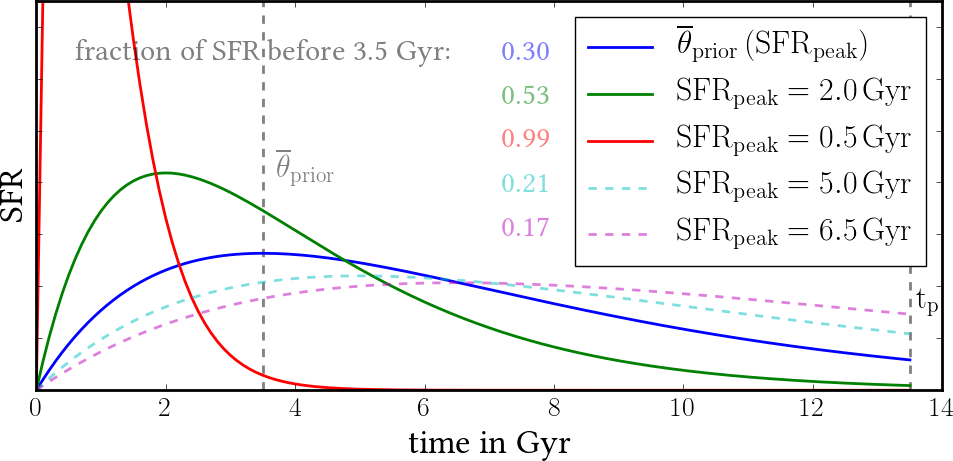}}
\caption{Illustration of the model parameter $\vec{\theta}_{5}=\mathrm{SFR}_\mathrm{peak}$, the epoch of the peak of the star formation rate (SFR), shown here in arbitrary units. As functional form for the SFR, we use the gamma distribution with $\mathrm{k}=2$ (see Equation\,\ref{eq:SFR}). Variations in the epoch of peak star formation of $\pm1,2\sigma_\mathrm{prior}$ from $\overline{\theta}_\mathrm{prior}$ are shown (see Table\,\ref{tab:abbreviations}). The fraction of stars being formed before 3.5\,Gyr is written next to the legend labels.}
\label{fig:SFR}
\end{figure}
To emulate the rise and fall of the SFR with time, \CP~ adopts a simple functional form, the gamma distribution:
\begin{equation}
\label{eq:SFR}
\mathrm{SFR}\left(t,k,\vartheta\right)=\frac{1}{\Gamma(k)\vartheta^k}t^{k-1}\exp\left(\frac{-t}{\vartheta}\right), \mathrm{for}\,k=2 \rightarrow \vartheta = \mathrm{SFR}_\mathrm{peak}. 
\end{equation}
 We fix the shape parameter $k=2$ such that the scale parameter ($\vartheta$) determines the peak of the SFR. This makes $\vec{\theta}_5=\mathrm{SFR}_\mathrm{peak}(=\vartheta)$ \CP 's fifth free parameter. The default distribution is depicted in Figure\,\ref{fig:SFR} together with the distributions resulting from $\sigma_\mathrm{prior}$ deviations, showing that this parametrization is highly non-linear. Still we chose this parametrization and prior distribution in order to obtain a smooth SFR, peaking early as observed in $L^\star$ galaxies \citep[fig\,4b]{VanDokkum2013b}. Whether the SFR should emulate the total SFR of the Milky Way, or the one near the Solar radius can be debated, and explored with \CP. Consequently we use an unnormalized SFR only being interested in the relative change of the SFR with time.

\begin{figure}
\resizebox{\hsize}{!}{\includegraphics{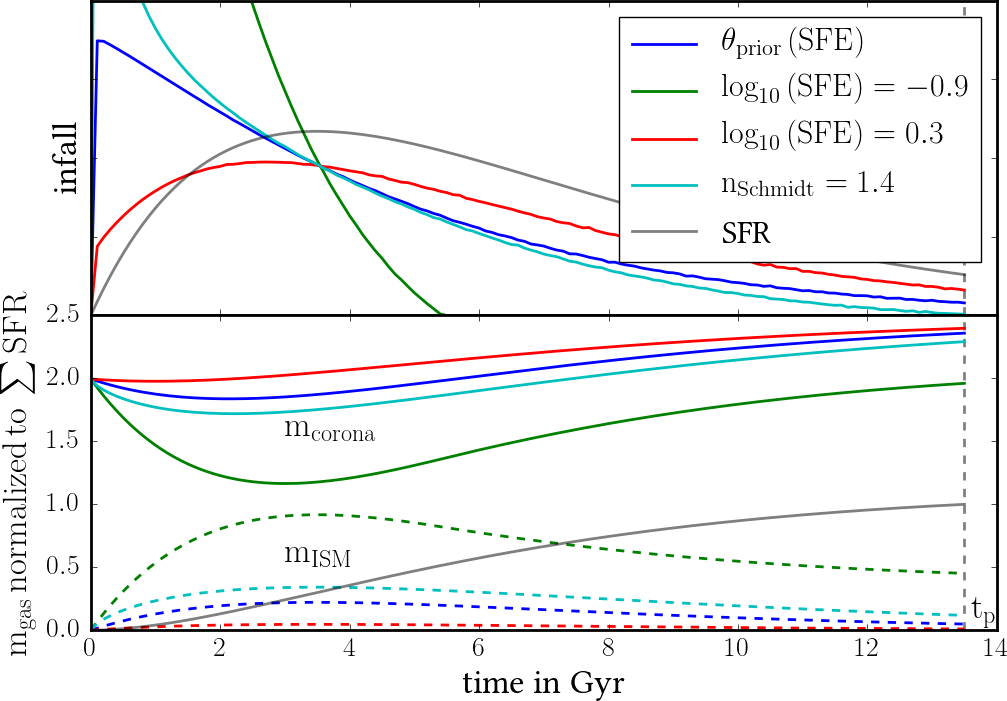}}
\caption{Illustration of how the gas infall and the gas masses (of ISM and corona) are governed by the model parameter "star formation efficiency", $\vec{\theta}_{4}=\mathrm{SFE}$. The upper panel shows the infall over time for $\overline{\theta}_\mathrm{prior}$ and $\pm2\sigma_\mathrm{prior}$ deviations. For comparison, a model with $\overline{\theta}_\mathrm{prior}$ but a Schmidt law exponent of 1.4 instead of our default 1.0 is shown (then the SFE parameter is no longer equal the star formation efficiency). In the lower panel gas mass of the corona (solid lines) and the ISM (dashed lines) are depicted for the four different cases from above, normalized to $\mathrm{m}_\mathrm{SFR}^\mathrm{tot}$. The SFR and its cumulative version are plotted in the upper and lower panel, respectively.}
\label{fig:SFE}
\end{figure}
At the beginning of a \CP\ run the ISM starts out with no mass, acquiring gas from the corona. In determining the gas infall needed to sustain a certain SFR, we generally assume a linear Schmidt law ($\mathrm{n}_\mathrm{Schmidt}=1$), for which our SFE-parameter literally is the star formation efficiency:
\begin{equation}
\mathrm{SFE}=\frac{\mathrm{m}_\mathrm{SFR}}{\mathrm{m}_\mathrm{ISM}^{\mathrm{n}_\mathrm{Schmidt}=1}}
\end{equation}
Using different power law exponents is possible and we illustrate the commonly used case of $\mathrm{n}_\mathrm{Schmidt}=1.4$ in Figure\,\ref{fig:SFE}, where the dependence of the infall and the ISM/corona gas mass onto the SFE parameter can be inspected. We center our prior on the SFE at the value by \citet{Bigiel2008} $\overline{\theta}_\mathrm{prior,4}=0.5\,\mathrm{Gyr}^{-1}$ with a variance of a factor of two (see Table\,\ref{tab:abbreviations}).

 After the enrichment of the ISM by the stellar feedback material from the previous SSPs the next SSP generation is formed, reducing the mass of the ISM. Not all feedback material is returned to the ISM, as there is some outflow fraction ($\mathrm{x}_\mathrm{out}=\vec{\theta}_6$. The outflow fraction, which is added to the corona, can be varied per process but we apply the same to all three enrichment processes for the sake of simplicity. Note that this parametrization is different from the commonly used mass-loading factor. As there are no meaningful observational constraints, we use a relatively broad prior peaking at an outflow fraction of $\overline{\theta}_\mathrm{prior,6}=0.5$.

\begin{figure}
\resizebox{\hsize}{!}{\includegraphics{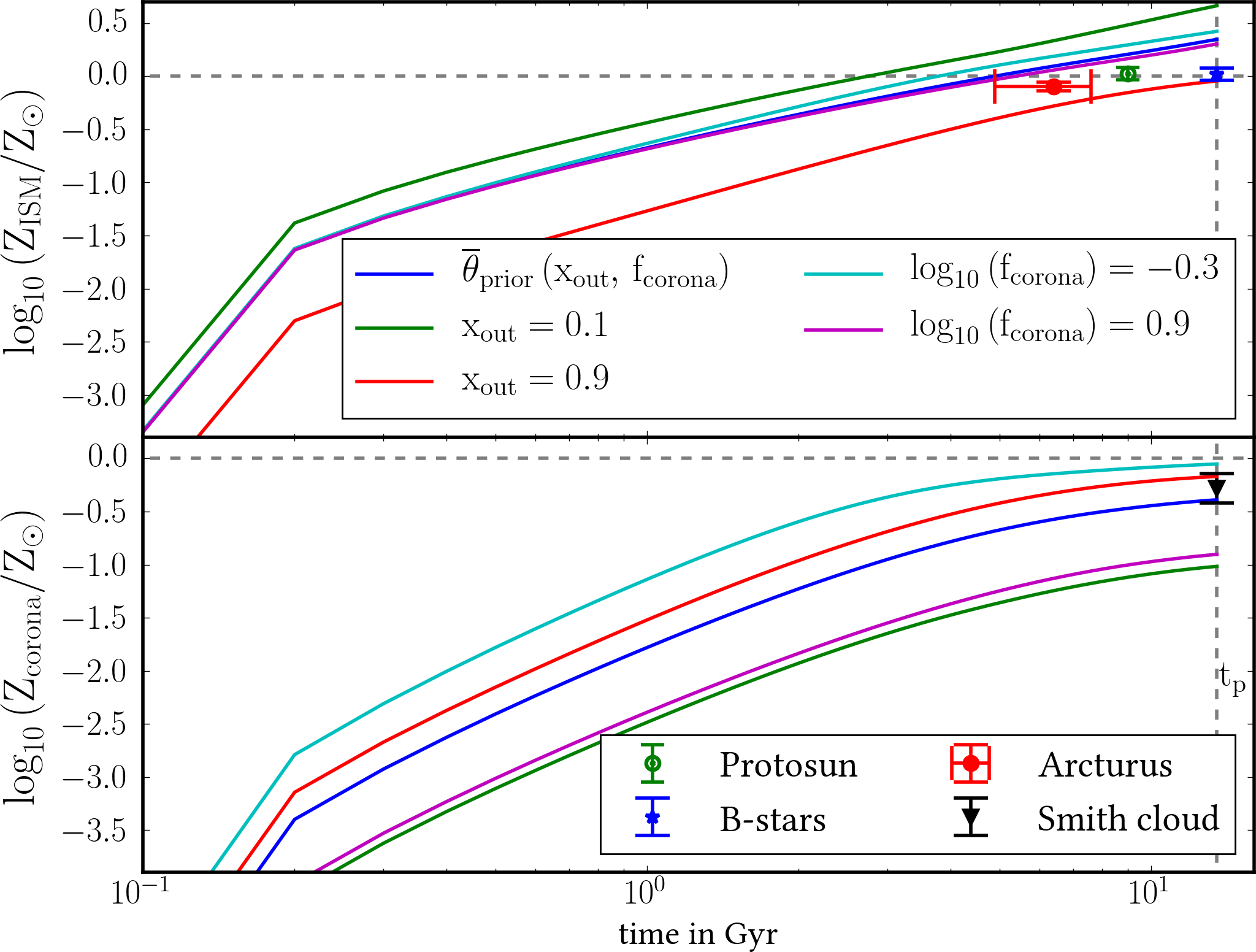}}
\caption{Illustrating the impact of model parameters $\vec{\theta}_{6,7}=\mathrm{x}_\mathrm{out},\mathrm{f}_\mathrm{corona}$, the mass loading or outflow fraction and the corona mass factor on the ISM metallicity. The resulting metallicity of ISM gas (upper panel) and corona gas (lower panel) is shown over time assuming the default yields. The results for $\overline{\theta}_\mathrm{prior}$ are compared with $\pm2\sigma_\mathrm{prior}$ deviations and our observational constraints.}
\label{fig:xout}
\end{figure}

The corona gas starts out with primordial abundances and its initial mass equals
\begin{equation}
\mathrm{m}_\mathrm{corona}\left(t=0\right)= \mathrm{f}_\mathrm{corona}\times \mathrm{m}_\mathrm{SFR}^{\mathrm{tot}},
\end{equation}
where
\begin{equation}
\mathrm{m}_\mathrm{SFR}^{\mathrm{tot}} := \Delta t \times \left(\sum_{t=0}^{\mathrm{t}_\mathrm{p}}\mathrm{m}_\mathrm{SFR}(t)\right)
\end{equation}
is the total stellar mass formed over the \CP\ run, and the "corona mass factor" is the last free parameter ($\mathrm{f}_\mathrm{corona}=\vec{\theta}_7$). Since the SFR is unnormalized and could represent e.g. the Solar birth chemical evolution zone, the corona gas can usually not be identified with a galactic halo (only if we were modeling whole galaxies as a Chempy single-zone, which could approximate the evolution of dwarf galaxies). Instead we must think of the corona as a gas reservoir, surrounding the chemical zone which we are modeling, and diluting its outflows. Therefore an outer thin disk zone would probably have a larger surrounding gas reservoir than an inner disk zone.

The corona is replenished with primordial gas (cosmic inflow) at the rate of the SFR; it loses mass to the ISM and it is chemically enriched by the outflowing fraction of the feedback material from the stellar component. Observational evidence from analysis of $L^\star$ galaxies \citep{Stern2016,Werk2014} have shown that the corona to stellar mass ratio at present-day ranges from 0.5 to 2. Because $\mathrm{f}_\mathrm{corona}$ is not directly identifiable with those data we chose a relatively broad prior centred around two with a variance of a factor of two (see Table\,\ref{tab:abbreviations}). In \CP, the corona is a simplified gas-reservoir surrounding the ISM, which is not available for star-formation. It could also be interpreted as a hot gas phase that is slowly cooling down, but we use the corona gas terminology throughout.

The effect of the parameters
$\mathrm{x}_\mathrm{out}$ and $\mathrm{f}_\mathrm{corona}$ is shown in Figure\,\ref{fig:xout}. The default model is compared to the 2$\sigma_\mathrm{prior}$ deviations in each parameter and data points from observations are included as well.

The four above steps are iterated over the course of 13.5\,Gyr ($\mathrm{t}_\mathrm{p}$) over a number of equidistant time-steps. We use 136 here (time resolution of $\Delta t = 0.1$~Gyr) which proves sufficient (even 28 time-steps yield similar results).

Since \CP~is very flexible, other mass flow as well as feedback prescriptions can be tested and more free parameters can easily be included (e.g. $k$, the shape parameter of the SFR or $\mathrm{n}_\mathrm{Schmidt}$). Tests were made and a reasonable set of parameters chosen so that the most important parameters should be included without over-fitting the problem. Other prescriptions and parameters can be tested by the interested reader using a documented user-case from the \CP~github repository.
\section{Observational constraints}
\label{ch:data}
We want to constrain the parameters of chemical evolution in the Galaxy by comparing \CP 's synthesized output (predictions) to data. It is not trivial to chose 
{\it a priori} the observations that are best suited to constrain the \CP 's model parameters. In the following we describe our fiducial set of observational evidence;
this is in some sense a set of minimal data, geared at demonstrating which aspects of abundance measurements matter most.

\begin{table*}
\begin{minipage}{\textwidth}
\begin{center}
\caption{Observations, ${\vec{\mathcal{O}}}$, used to constrain the \CP~ parameters.}
\begin{tabular}{ l|c| c }
Abbreviation & Nomenclature & Description and source\\
  \hline
   \multicolumn{3}{c}{Stellar abundances}\\
   \hline
  $\{1\}:\vec{[\mathrm{X}/\mathrm{H}]}_{\odot,\mathrm{birth}}$ & Sun & protosolar abundances from \citet{Asplund2009} \\
  $\{2\}:\vec{[\mathrm{X}/\mathrm{H}]}_\mathrm{B-stars}$ & B-stars & nearby B-stars by \citet{Nieva2012} as ISM proxy \\
  $\{3\}:\vec{[\mathrm{X}/\mathrm{Fe}]}_\mathrm{Arc}$ & Arcturus & abundances and age as derived by \citet{Ramirez2011}\\
   \hline
   \multicolumn{3}{c}{Additional constraints\footnote{We combine each stellar abundance with the SN-ratio and corona metallicity in an observational set, e.g. for the Sun: \{1,4,5\} and call it "Sun+" indicating the additional constraints (which we add in order to reproduce Milky Way-like global constraints in each zone)} ("+")}\\
   \hline
  $\{4\}:\mathrm{CC}/\mathrm{Ia}$ & SN-ratio & CC-SN to SN\,Ia ratio in Sbc/d galaxies at present day \citep{Mannucci2005}\\
  $\{5\}:\mathrm{Z}_{\mathrm{Smith}}$ & corona metallicity & metallicity of Smith cloud measured by \citet{Fox2016}
  \label{tab:obs_constraints}
\end{tabular}
\end{center}

\end{minipage}
\end{table*}
\subsection{Solar abundances}
Arguably, the most accurate and precise stellar element abundances are the present-day Solar photospheric abundances, $\vec{[\mathrm{X}/\mathrm{H}]}_\odot$, by \cite[tab.\,1]{Asplund2009}. Still the abundance determination from spectroscopy relies on theoretical models and their systematic error is probably underestimated \citep{Jofre2016}, as illustrated by the ongoing debate regarding the Solar Oxygen abundance \citep{Steffen2015}. But the Solar abundances has been verified using meteorites \citep{Lodders2009}, other Solar system bodies elemental abundances \citep{Lawler1989,McDonough1995} as well as helioseismological inference \citep{Basu2004}. At the same time the age of the Sun is very well constrained \citep[e.g.][]{Dziembowski1998}, hence we can map the \CP~ISM abundances from $\sim$4.5\,Gyr ago $\left(\vec{[\mathrm{X}/\mathrm{H}]}_\mathrm{ISM}(\mathrm{t}_\mathrm{p}-4.5)=\vec{[\mathrm{X}/\mathrm{H}]}_\mathrm{SSP}(\mathrm{t}_\mathrm{p}-\tau_\odot)\right)$ onto the Solar birth abundances (protosolar), $\vec{[\mathrm{X}/\mathrm{H}]}_{\odot,\mathrm{birth}}$. We adopt protosolar abundances by adding 0.04\,dex (0.05\,dex for He) to the photospheric abundances which is the depletion of heavy elements due to diffusion processes \citep{Turcotte2002} in the Sun; and we add 0.01\,dex to the abundance uncertainty, accounting for this imperfect correction.

This leaves the decision of how many elements one can sensibly include in \CP 's prediction and data comparison. For our analysis we consider the elements up to Ni (for both yield sets CC-SN and SN\,Ia tables provide all those elements and the AGB tables provide elements at least up to Si). 
We exclude Li, Be and B because other nucleosynthetic channels (i.e. cosmic ray spallation and nova outbursts), which we do not model, contribute substantially to the enrichment of these elements \citep{Reeves1970, Romano1999}. Similarly Li can easily be destroyed in the Solar photosphere altering its abundance compared to the initial Solar composition \citep{Asplund2009}. We also excluded Cl and Sc because the mismatch between predictions and observations was too large \footnote{This is of course an arbitrary choice  and biases our results (though the bias is small because we use many elements). Conceptually, it would seem attractive to include all available elements but that would require us to model the effects changing photospheric elemental abundances and we would need to include all nucleosynthetic channels and have uncertainties on the yield tables as well.}. We use all remaining elements for our analysis: He, C, N, O, F, Ne, Na, Mg, Al, Si, P, S, Ar, K, Ca, Ti, V, Cr, Mn, Fe, Co and Ni. Changing the Solar abundances to \citet[tab.\,6]{Lodders2009}, which are based on meteoritic data, does not affect our results significantly.
\subsection{The present-day ISM Abundances, with early B-stars as Proxy}
The present-day ISM abundances are a crucial anchor for GCE models. One of the best measurements is \citet[tab.\,9]{Nieva2012}, who determined He, C, N, O, Ne, Mg, Si and Fe abundances for 20 nearby (within 0.5\,kpc) early B-stars, with a star-to-star scatter comparable and often lower than the Solar abundance precision of \cite{Asplund2009}. This suggests that the present-day ISM is well mixed and establishes a "cosmic abundance standard", $\vec{[\mathrm{X}/\mathrm{H}]}_\mathrm{B-stars}$, which we use to constrain the ISM abundance at the end of the simulation, $\vec{[\mathrm{X}/\mathrm{H}]}_{\mathrm{ISM}(\mathrm{t}_\mathrm{p})}$.    
\subsection{Arcturus as the best studied  $\alpha$-enhanced star}
Arcturus is a well-studied giant star with $\alpha$-enhanced abundances and an age of about 7\,Gyr. \citet{Ramirez2011} find an $\mathrm{[Fe/H]}$ of $-0.52\pm0.04$\,dex and provide abundances for C, O, Na, Mg, Al, Si, K, Ca, Ti, V, Cr, Mn, Co and Ni, in common with our elemental choice for the Sun. We exclude C, since the photospheric value in a giant does not represent the initial abundance because of dredge-up \citep{Iben1965}. We do not apply any "proto-Arcturus" correction to its present-day photospheric abundances, $\vec{[\mathrm{X/Fe}]}_\mathrm{Arcturus}$, which we assume to reflect the ISM composition at the time of its birth, $\vec{[\mathrm{X}/\mathrm{Fe}]}_\mathrm{ISM}(\mathrm{t}_\mathrm{p}-\tau_\mathrm{Arcturus})$. Arcturus' age of 7.1\,Gyr$_{-1.2}^{+1.5}$ by \citet{Ramirez2011} is less well known than the Sun's but that uncertainty turns out not to affect our results significantly.
\subsection{Observed incidence of supernovae}
A more global observational constraint is the ratio of CC-SNe to SNe\,Ia, $\mathrm{CC}/\mathrm{Ia}$. The SN-ratio contains information about the SFR (since CC-SNe trace the SFR directly and SNe\,Ia with a delay), about the IMF (number of CC-SN) and the number of SN\,Ia exploding per Solar mass. Since data for the Milky Way are not available, we use Sbc/d galaxies from \citet{Mannucci2005}. For SN~Ia they measure $0.17^{+0.068}_{-0.063}$ and $0.86^{+0.384}_{-0.359}$ for CC-SN. This corresponds to a ratio of $5.06^{+6.57}_{-2.95}$ for which we use $\log_{10}\left(\mathrm{CC}/\mathrm{Ia}(\mathrm{t}_\mathrm{p})\right)=0.7\pm0.37$, simplifying to a Gaussian error model.
\subsection{Corona metallicity}
Observational constraints on the abundances of the corona gas turn out to be important to reduce parameter covariances (or even degeneracies) in 
\CP . We know from \citep{Smith1963,Fox2016} that the material, falling onto the Galactic disk, is enriched, not pristine.
As observational constraint for the present-day corona gas metallicity we use the "Smith cloud" value of about half Solar: $\log_{10}\left(\mathrm{Z}_{\mathrm{Smith\,cloud}}\right)=-0.28\pm0.14\,\mathrm{Z}_\odot$. Recent reanalysis of observational data by \citet[Fig.\,3]{Stern2016} fitting individual metallicities in the circum-galactic medium (CGM) of $L^*$-galaxies in the COS-halos survey yields similar corona abundances with a range of $\log_{10}\left(\mathrm{Z}_{\mathrm{CGM}}\right)=-0.2\pm0.3\,\mathrm{Z}_\odot$.
\subsection{Combination of observational data}
As observational constraints we will usually use a combination of stellar abundance, corona metallicity and SN-ratio and indicate this by adding a "+" to the star's name (e.g. Sun+ for \{1,4,5\}, cf. Table\,\ref{tab:obs_constraints}). The reason for adding the latter two constraints (which are, strictly speaking, not "Milky Way data") is that we demand our model to reproduce a few basic, globally observed properties of $L^\star$ galaxies; they also constrain the otherwise ill-determined and degenerate ISM parameters $\mathrm{x}_\mathrm{out}$ and $\mathrm{f}_\mathrm{corona}$.
\subsection{Comparative data set: APOGEE giants}
In order to bring our predictions into perspective with Galactic stellar populations we use an APOGEE \citep{Majewski2016} giant sample with DR13 ASPCAP \citep{SDSSCollaboration2016} abundances, for which ages have been derived from C/N ratios \citep{Ness2016}. We only chose stars for which the \citet{Ness2016} $\chi^2$ fit is better than 0.9, leaving us with a sample of $\approx$$20,000$ stars to which we compare our resulting chemical evolution tracks in Figure\,\ref{fig:apogee}.

Another important prediction is the metallicity distribution function (MDF). We will compare Chempy predictions with the APOGEE DR13 red-clump catalogue \citep{Bovy2014b} for which good distance estimates are available.

\subsection{Omitted observational constraints}
\label{ch:MDF}
We are not using the APOGEE data as constraints for our inference because e.g. the \CP~Solar zone (the chemical enrichment of the ISM leading to the abundances of the Sun) can not be identified with the ISM evolution at the Solar neighbourhood (the gas which is present here now could have experienced chemical enrichment at different places in the Galaxy). Vice versa a stellar sample from the Solar neighbourhood will also not be representative for the ISM evolution at Solar Galactic radius because the stellar birth radius is not preserved.

Similarly we are not using present-day stellar or gas densities as observational constraints. The \CP~SFR can always be renormalized to match a specific stellar density. But since the ISM gas mass is tight to the SFR via the SFE a constraint on the gas density will force a specific present-day SFR value, which will bias our SFR$_\mathrm{peak}$ inference, even more so, since we are using a very simple SFR parametrization.
\section{Constraining parameters via Bayesian inference}
\label{ch:method}
As depicted in Figure\,\ref{fig:time_integration} a single \CP~run evaluates the unnormalised posterior PDF for a specified set of observations $\vec{\mathcal{O}}_\mathrm{s}\subseteq\mathcal{O}$, at a specific point $\vec{\theta}$ in parameter space (steps 1 - 5). The complete posterior PDF can be approximated using \CP\ within an MCMC scheme (step 6). The steps are as follows: 
\begin{enumerate}
\item A point in parameter space ($\vec{\theta}$) is chosen. This sets the log prior:
\begin{equation}
\ln\left(\frac{\mathcal{P}}{\mathcal{P}_\mathrm{max}}\right) = -\sum_i^{\mathrm{n}_{\vec{\theta}}}\frac{\left(\theta_i-\overline{\theta}_{\mathrm{prior}_i}\right)^2}{2\sigma_{\mathrm{prior}_i}^2},
\end{equation}
normalized by its maximum $\mathcal{P}_\mathrm{max}$;
$\mathrm{n}_{\vec{\theta}}$ is the number of free parameters (i.e. 7, the dimensions of $\vec{\theta}$). The analytic values for the prior mean ($\overline{\theta}_{\mathrm{prior}_i}$) and the standard deviation ($\sigma_{\mathrm{prior}_i}$) of the prior distribution are given in Table\,\ref{tab:abbreviations}. Because of the imposed limits on the parameter values the real prior distribution is slightly distorted which has a negligible effect as can be seen in the first and second row of Table\,\ref{tab:results_small}.

\item \CP~then integrates the system state, $\vec{\mathrm{S}}_{\vec{\theta}}(t)$, 
from the initial condition
 ($\vec{\mathrm{S}}_{\vec{\theta}}(t=0)$) 
 to the final epoch, usually the present time at $13.5$~Gyrs  ($\vec{\mathrm{S}}_{\vec{\theta}}(t=\mathrm{t}_\mathrm{p})$) in steps of $\Delta t=0.1\,\mathrm{Gyr}$.
The system state keeps track of the chemical composition and mass of each component:
\begin{equation}
\mathrm{Chempy}(\vec{\theta}) = \vec{\mathrm{S}}_{\vec{\theta}}(t)=\left\{\mathrm{m}_{\mathrm{ISM,corona,CSP}}(t),\vec{[\mathrm{X/H}]}_\mathrm{ISM,corona,CSP}(t)  \right\}.
\end{equation}
Additionally derived quantities from stellar evolution (e.g. feedback per process, mass in remnants, number of feedback events) are saved for the stellar component (cf. Figure\,\ref{fig:mass-flow}).

\item The previously chosen set of observations ($\vec{\mathcal{O}}_\mathrm{s}$) is compared to the corresponding predictions ($\vec{\mathrm{d}}_\mathrm{s}(\vec{\theta})$) which are derived from the system state $\left(\vec{\mathrm{S}}_{\vec{\theta}}(t)\right)$,
\begin{equation}
\vec{\mathrm{S}}_{\vec{\theta}}(t)\rightarrow\vec{\mathrm{d}}_\mathrm{s}(\vec{\theta})\subseteq
\left(
\begin{matrix}
\vec{[\mathrm{X}/\mathrm{H}]}_\mathrm{SSP}(\mathrm{t}_\mathrm{p}-\tau_\odot)\\
\vec{[\mathrm{X}/\mathrm{H}]}_\mathrm{ISM}(\mathrm{t}_\mathrm{p})\\
\vec{[\mathrm{X}/\mathrm{Fe}]}_\mathrm{SSP}(\mathrm{t}_\mathrm{p}-\tau_\mathrm{Arc})\\
\mathrm{CC}/\mathrm{Ia}(\mathrm{t}_\mathrm{p})\\
\mathrm{Z}_\mathrm{corona}(\mathrm{t}_\mathrm{p})
\end{matrix}
\right)
\leftrightarrow
\vec{\mathcal{O}}_\mathrm{s}\subseteq
\left(
\begin{matrix}
\vec{[\mathrm{X}/\mathrm{H}]}_{\odot,\mathrm{birth}}\\
\vec{[\mathrm{X}/\mathrm{H}]}_\mathrm{B-stars}\\
\vec{[\mathrm{X}/\mathrm{Fe}]}_\mathrm{Arc}\\
\mathrm{CC}/\mathrm{Ia}\\
\mathrm{Z}_\mathrm{Smith\,cloud}
\end{matrix}
\right).
\end{equation}
In our case the age of the star $\tau_\star$ is $4.5\,\mathrm{Gyr}$ for the Sun and $7.1\,\mathrm{Gyr}$ for Arcturus, and all other observations are compared to the predictions at the end of the simulation, $\mathrm{t}_\mathrm{p}=13.5\,$Gyr.

\item The (log) likelihood of the observational constraint given the model predictions $\mathcal{L}=\mathrm{P}\left(\vec{\mathcal{O}}_\mathrm{s}|\vec{\mathrm{d}}_\mathrm{s}(\vec{\theta})\right)$,  is also normalized to its maximum value ($\mathcal{L}_\mathrm{max}$) resulting in the log likelihood being written as 
\begin{equation}
\ln\left(\frac{\mathcal{L}}{\mathcal{L}_\mathrm{max}}\right) = -\sum_i^{\mathrm{n}_\mathrm{data\,points}}\frac{\left(\mathcal{O}_{\mathrm{s,}i}-\mathrm{d}_{\mathrm{s},i}\right)^2}{2\sigma_{\mathrm{obs}_i}^2}.
\end{equation}
The index $i$ goes over all data points within $\mathcal{O}_\mathrm{s}$, each with their associated standard deviation (reported observational error, $\sigma_{\mathrm{obs}_i}$).
For Sun+ $\mathrm{n}_\mathrm{data\,points}$ would be 24, i.e. 22 elemental abundances and one data point each for the SN-ratio and the corona metallicity.
\item The unnormalized log posterior ($\mathbb{P}$), i.e. the probability of the parameters given the data $\mathrm{P}(\vec{\theta}|\vec{\mathcal{O}}_\mathrm{s})=\mathrm{P}(\vec{\mathcal{O}}_\mathrm{s}|\vec{\mathrm{d}}_\mathrm{s}(\vec{\theta}))\mathrm{P}(\vec{\theta})$, is then:
\begin{equation}
\mathbb{P}(\vec{\theta}|\vec{\mathcal{O}}_\mathrm{s})= \ln\left(\frac{\mathcal{L}}{\mathcal{L}_\mathrm{max}}\right) + \ln\left(\frac{\mathcal{P}}{\mathcal{P}_\mathrm{max}}\right).
\end{equation}
In essence the posterior is a product of normal distributions, with $\mathrm{n}_{\vec{\theta}}=7$ terms from the prior and $\mathrm{n}_\mathrm{data\,points}$ (depending on the set of observations used) terms from the likelihood.

For clarification, the log posterior $\mathbb{P}(\vec{\theta}|\vec{\mathcal{O}}_\mathrm{s})$
values with these definitions are:
\begin{itemize}
\item $0$, if parameters are at their peak values ($\vec{\theta}=\overline{\vec{\theta}}_\mathrm{prior}$) and the predictions reproduce the data perfectly ($\vec{\mathrm{d}}_\mathrm{s}=\vec{\mathcal{O}}_\mathrm{s}$).
\item $-0.5\,(-2, -4.5, -8)$ if only one parameter or one data point would be one (two, three, four) standard deviation ($\sigma_\mathrm{prior},\sigma_\mathrm{\vec{\mathrm{obs}}}$) away from its maximum value ($\overline{\theta}_\mathrm{prior},\vec{\mathcal{O}}_\mathrm{s}$).   
\item $-0.5\times(\mathrm{n}+\mathrm{m})$ if n parameter and m data points would be one standard deviation away from its default.
\end{itemize}

\item A single evaluation of the posterior function ("one \CP~run") with a specific set of observations
\begin{equation}
\mathrm{Chempy}_{\mathcal{O}_\mathrm{s}}(\vec{\theta})\rightarrow\mathbb{P}(\vec{\theta}|\vec{\mathcal{O}}_\mathrm{s})
\end{equation}
requires a few seconds on a modern CPU. Since we are interested in the complete posterior PDF of the \CP~parameters, we use MCMC sampling to approximate it. We employ {\tt emcee} \citep{Foreman2013}, which can easily parallelize the process. For each MCMC run we initialise 64\,walkers in a small cloud around $\overline{\vec{\theta}}_\mathrm{prior}$. In each iteration each walker evaluates the posterior function at a new position and rejects or accepts it depending on how the new posterior compares to the posterior of the last accepted evaluation. 
 After a burn-in phase, the walkers then actually sample the PDF. We check whether the mean posterior of those walkers converge (which usually happens after 100 burn-in iterations) and leave the MCMC chain stabilize for 200 more steps. The parameter distribution of the 5000 last entries, $j$, of the flattened MCMC chain
\begin{equation}
\left\{\mathrm{Chempy}_{\mathcal{O}_\mathrm{s}}(\vec{\theta}_j,\vec{\lambda})\right\}_\mathrm{stabilized}\rightarrow\left\{\vec{\theta}_{j,\vec{\mathcal{O}_\mathrm{s}},\vec{\lambda}}\right\}=:\vec{\theta}_j\left(\mathcal{O}_\mathrm{s},\lambda\right)\approx\mathrm{P}(\vec{\theta}|\vec{\mathcal{O}}_\mathrm{s},\vec{\lambda}),
\end{equation}
is used as a representation for the posterior PDF over the parameter space. The hyperparameters $\lambda$ are now written explicitly, because for each subset of $\mathcal{O}$ for which we will do our inference we will also derive parameter distributions when using \CP~with an alternative set of yields.  
\end{enumerate}
An example of the posterior PDF for \CP~used with Sun+ as observational constraint and the default yield set, i.e. $\vec{\theta}_j\left(\mathrm{Sun+},\mathrm{default}\right)$, can be seen in Figure\,\ref{fig:parameter_space}, together with the prior distribution in dashed lines. One such MCMC run takes approximately 1\,hour on a 64\,core machine, which allows us to extensively test our results.
\section{Results}
\label{ch:results}

We will now present and discuss the inferred \CP~parameters (Table\,\ref{tab:abbreviations}), when using \CP~with different subsets of our observations (\,\ref{tab:obs_constraints}) and with the two different yield sets (Table\,\ref{tab:yield}). 

\begin{table*}[]
\begin{tiny}
\centering
\caption{Inferred parameters (16, 50, 84 percentiles of $\theta_\mathrm{posterior}$), for different observational subsets ($\mathcal{O}_\mathrm{s}$), and different yield sets (default, alternative).}
\label{tab:results_small}
\begin{tabular}{l|c||cc|cc|cc|cc|cc|cc|cc}
\hline
observational set & posterior & \multicolumn{2}{c|}{$\alpha_\mathrm{IMF}$} & \multicolumn{2}{c|}{$\log_{10}\left(\mathrm{N}_\mathrm{Ia}\right)$} & \multicolumn{2}{c|}{$\log_{10}\left(\tau_\mathrm{Ia}\right)$} & \multicolumn{2}{c|}{$\log_{10}\left(\mathrm{SFE}\right)$} & \multicolumn{2}{c|}{SFR$_\mathrm{peak}$} & \multicolumn{2}{c|}{x$_\mathrm{out}$} & \multicolumn{2}{c}{$\log_{10}\left(\mathrm{f}_\mathrm{corona}\right)$} \\
$\vec{\mathcal{O}}_\mathrm{s}$ &   $\mathbb{P}_\mathrm{max}$  & $\mu$ & $\sigma$ & $\mu$ & $\sigma$& $\mu$ & $\sigma$& $\mu$ & $\sigma$& $\mu$ & $\sigma$& $\mu$ & $\sigma$& $\mu$ & $\sigma$\\

\hline 
\{\} (analytic prior) &    0     & -2.29 & $\pm0.20$ & -2.75 & $\pm0.30$ & -0.80 & $\pm0.30$ & -0.30 & $\pm0.30$ & 3.50 & $\pm1.50$ & 0.50 & $\pm0.20$ & 0.30  & $\pm0.30$ \\
\hline
\{\} (prior-only run) & -0.1   & -2.29 & $\pm0.20$ & -2.77 & $^{+0.29}_{-0.30}$ & -0.80 & $\pm0.30$ & -0.29 & $^{+0.31}_{-0.30}$ & 3.47 & $^{+1.47}_{-1.50}$ & 0.51 & $\pm0.20$ & 0.30  & $\pm0.30$ \\
\hline
\multicolumn{16}{c}{synthetic observations (produced with default yield set and $\overline{\theta}_\mathrm{prior}$ parameters, Section\,\ref{ch:mock} Figure\,\ref{fig:mock_default})}\\
\hline
Sun+ (default yield) & -5.5   & -2.28 & $^{+0.08}_{-0.06}$ & -2.74 & $^{+0.10}_{-0.11}$ & -0.81 & $^{+0.30}_{-0.28}$ & -0.25 & $\pm0.23$ & 3.79 & $^{+1.18}_{-1.07}$ & 0.49 & $\pm0.12$ & 0.26  & $\pm0.24$ \\
\hline
Sun+ (alternative)& -34.4   & -2.35 & $^{+0.07}_{-0.05}$ & -3.10 & $^{+0.11}_{-0.12}$ & -0.83 & $^{+0.29}_{-0.30}$ & 0.15 & $^{+0.25}_{-0.23}$ & 2.14 & $^{+0.43}_{-0.25}$ & 0.52 & $^{+0.08}_{-0.10}$ & 0.18  & $^{+0.23}_{-0.24}$ \\
\hline
\multicolumn{16}{c}{single-zone (Section\,\ref{ch:separate_zones}, Figure\,\ref{fig:multiple_obs}) for default yield set}\\
\hline
Sun+ (fig.\,\ref{fig:parameter_space}) & -105.8   & -2.46 & $\pm0.04$ & -3.07 & $_{-0.11}^{+0.10}$ & -0.80 & $_{-0.30}^{+0.29}$ & -0.31 & $_{-0.15}^{+0.14}$ & 3.02 & $_{-0.61}^{+0.78}$ & 0.47 & $\pm0.09$ & -0.11 & $^{+0.22}_{-0.23}$ \\
\hline
B-stars+ & -22.3   & -2.43 & $\pm0.05$ & -2.92 & $^{+0.10}_{-0.12}$ & -0.81 & $\pm0.30$ & -0.49 & $^{+0.21}_{-0.19}$ & 3.75 & $^{+1.22}_{-1.05}$ & 0.68 & $^{+0.10}_{-0.11}$ & 0.15  & $^{+0.18}_{-0.19}$ \\
\hline
Arcturus+ & -234.4   & -2.32 & $^{+0.07}_{-0.06}$ & -3.55 & $^{+0.15}_{-0.14}$ & -0.61 & $^{+0.35}_{-0.33}$ & -0.63 & $^{+0.22}_{-0.18}$ & 2.77 & $^{+1.22}_{-0.79}$ & 0.63 & $^{+0.14}_{-0.16}$ & 0.17  & $\pm0.19$\\
\hline
Sun+,B-stars,Arcturus & -405.9   & -2.40 & $\pm0.04$ & -3.87 & $^{+0.09}_{-0.07}$ & 0.16 & $^{+0.21}_{-0.27}$ & -0.09 & $\pm0.02$ & 1.22 & $^{+0.04}_{-0.05}$ & 0.80 & $\pm0.03$ & 0.42  & $\pm0.12$ \\
\hline
\multicolumn{16}{c}{alternative yield set}\\
\hline
Sun+ & -87.4   & -2.51 & $_{-0.03}^{+0.04}$ & -3.49 & $_{-0.12}^{+0.10}$ & -0.88 & $_{-0.29}^{+0.26}$ & 0.11 & $_{-0.18}^{+0.23}$ & 2.14 & $_{-0.21}^{+0.23}$ & 0.44 & $^{+0.07}_{-0.06}$ & -0.11 & $^{+0.17}_{-0.19}$ \\
\hline
B-stars+ & -14.9   & -2.31 & $^{+0.07}_{-0.06}$ & -2.87 & $^{+0.10}_{-0.11}$ & -0.79 & $^{+0.30}_{-0.29}$ & -0.36 & $^{+0.35}_{-0.30}$ & 4.47 & $\pm1.16$ & 0.74 & $^{+0.07}_{-0.12}$ & 0.31  & $\pm0.17$ \\
\hline
Arcturus+ & -209.0   & -2.19 & $\pm0.07$ & -3.08 & $^{+0.31}_{-0.64}$ & 0.74 & $^{+0.09}_{-1.13}$ & -0.23 & $^{+0.25}_{-0.36}$ & 4.61 & $^{+1.08}_{-1.06}$ & 0.75 & $^{+0.07}_{-0.11}$ & 0.44  & $^{+0.13}_{-0.15}$\\
\hline
Sun+,B-stars,Arcturus & -395.2   & -2.39 & $^{+0.04}_{-0.03}$ & -3.22 & $^{+0.16}_{-0.21}$ & 0.70 & $^{+0.03}_{-0.05}$ & 0.16 & $^{+0.11}_{-0.10}$ & 3.20 & $^{+1.14}_{-0.58}$ & 0.74 & $^{+0.10}_{-0.13}$ & -0.33  & $\pm0.19$ \\
\hline
\hline
\multicolumn{2}{c||}{multi-zone scheme (Section\,\ref{ch:multizone})} & \multicolumn{6}{c|}{mutual SSP parameters} & \multicolumn{8}{c}{individual ISM parameters}\\
\hline
Sun+,  &    &  &  &  &  &  &  & -0.17 & $\pm0.15$ & 2.65 & $^{+0.47}_{-0.46}$ & 0.54 & $\pm0.08$ & 0.03  & $^{+0.18}_{-0.16}$ \\
B-stars+, &    &  &  &  &  &  &  & -0.38 & $^{+0.22}_{-0.16}$ & 3.54 & $^{+1.04}_{-0.80}$ & 0.67 & $^{+0.10}_{-0.09}$ & 0.19  & $^{+0.14}_{-0.16}$ \\
Arcturus+ (default) &  \multirow{-3}{*}{-371.4}  & \multirow{-3}{*}{-2.44} & \multirow{-3}{*}{$\pm0.03$} & \multirow{-3}{*}{-2.99} & \multirow{-3}{*}{$^{+0.15}_{-0.20}$} & \multirow{-3}{*}{0.49} & \multirow{-3}{*}{$^{+0.06}_{-0.10}$} & -0.54 & $^{+0.14}_{-0.15}$ & 4.42 & $^{+1.22}_{-0.90}$ & 0.50 & $^{+0.14}_{-0.11}$ & -0.19  & $\pm0.17$ \\
\hline
Sun+,  &    &  &  &  &  &  &  & 0.29 & $^{+0.20}_{-0.26}$ & 2.20 & $^{+0.46}_{-0.26}$ & 0.56 & $\pm0.07$ & 0.12  & $^{+0.20}_{-0.15}$ \\

B-stars+, &    &  &  &  &  &  &  & 0.01 & $^{+0.21}_{-0.27}$ & 3.30 & $^{+0.79}_{-0.57}$ & 0.71 & $^{+0.06}_{-0.07}$ & 0.19  & $^{+0.14}_{-0.15}$ \\
Arcturus+ (alternative) &  \multirow{-3}{*}{-326.2}  & \multirow{-3}{*}{-2.40} & \multirow{-3}{*}{$^{+0.04}_{-0.03}$} & \multirow{-3}{*}{-3.15} & \multirow{-3}{*}{$^{+0.18}_{-0.19}$} & \multirow{-3}{*}{0.78} & \multirow{-3}{*}{$^{+0.05}_{-0.04}$} & -0.02 & $\pm0.21$ & 5.12 & $\pm0.92$ & 0.61 & $^{+0.07}_{-0.08}$ & 0.14  & $\pm0.14$ \\
\hline

\end{tabular}
\end{tiny}
\end{table*}

\subsection{Test with synthesized observations}
\label{ch:mock}
\begin{figure*}
\resizebox{\hsize}{!}{\includegraphics{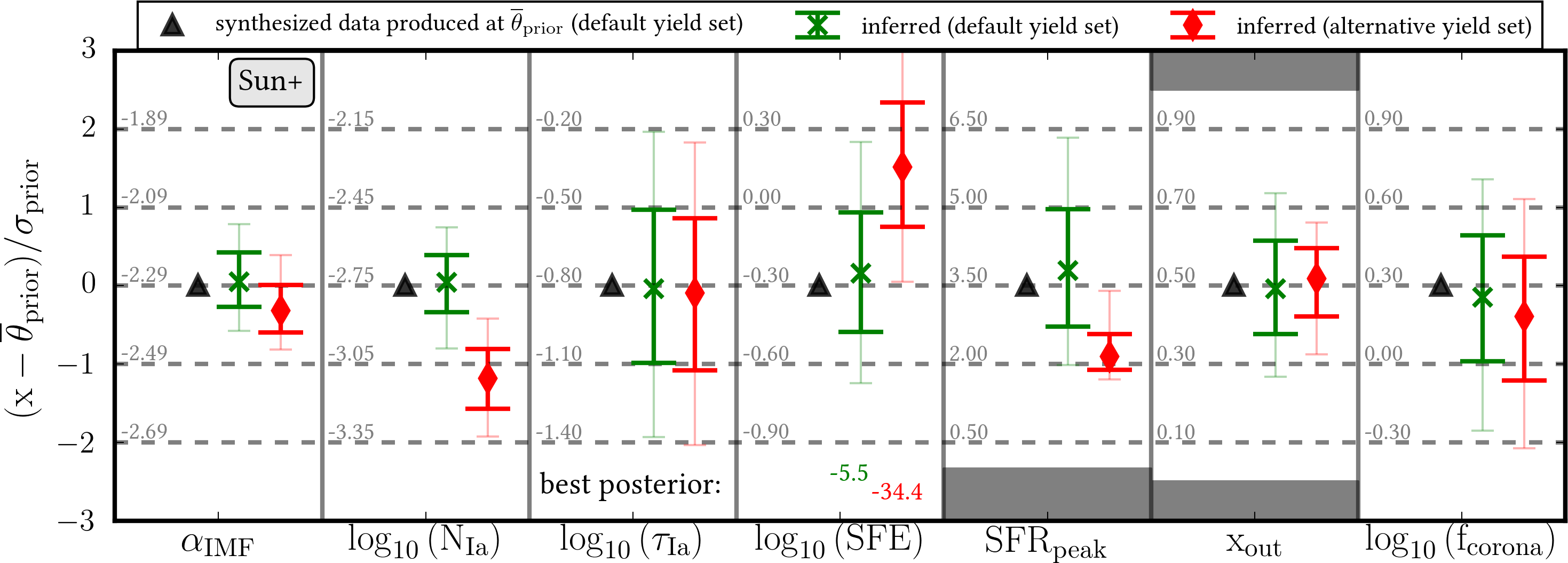}}
\caption{
llustration of model parameter constraints that can be excepted from observational constraints that are like Sun+ (Table\,\ref{tab:obs_constraints}). We generated Sun+ synthetic data from $\overline{\theta}_\mathrm{prior}$ parameters (black triangles) with the default yield set. We then generate 10 realizations of such Sun+-like synthetic data, with errors $\sigma_\mathrm{obs}$, and infer \CP~parameter PDFs for each, using the default (green) and the alternative (red) yield set. The marginalized parameter distributions are plotted on the prior scale, with the absolute parameter values noted on the left side of each box in grey. The median and the 15.9 \& 84.1 (2.3 \& 97.7) PDF percentiles are shown in solid (transparent) lines for each parameter. The maximum posterior value of all 10\,MCMC runs is given in the bottom of the SFE box. The grey area indicates limits of the parameter space. This Figure illustrates that the Sun+ abundances already provide constraints on $\alpha_\mathrm{IMF}$ and $\log_{10}(\mathrm{N}_\mathrm{Ia})$.}
\label{fig:mock_default}
\end{figure*}

We synthesize data for Sun+ observational constraints by taking \CP 's predictions as observations, assuming the default parameter configuration (see $\overline{\vec{\theta}}_\mathrm{prior}$ in Table\,\ref{tab:abbreviations}). We perturb the predictions by the observational uncertainties, $\sigma_\mathrm{obs}$, to obtain mock observations. Then we start a parameter inference (an MCMC run) on these synthetic data and see how well the \textit{injected} parameters, $\overline{\vec{\theta}}_\mathrm{prior}$, can be \textit{retrieved} and how large the uncertainty is. If we would not perturb the predictions by $\sigma_\mathrm{obs}$ the initial parameters are inferred almost perfectly. By perturbing the predictions by $\sigma_\mathrm{obs}$ we can estimate the internal error of our method for Sun+ as observational constraint. At the same time we test how well the parameters are inferred when using the \textit{alternative} yield set on synthesized data produced by the \textit{default} yield set. In Figure\,\ref{fig:mock_default} we show the combined parameter distributions of 10 such MCMC inferences (mock data creation and inference is repeated 10\,times), for each yield set.

Figure\,\ref{fig:mock_default} shows that the inferred parameter distributions (green crosses) are consistent with the input parameters (black triangles), if we use the default yield set (from which the mock data was synthesized). We can also see how well each parameter is constrained by the few observational constraints, essentially only the Sun's element abundances, the corona metallicity and the observed SN-ratio. In units of $\sigma_\mathrm{prior}$ the parameters' PDF variance is only 0.34 for $\alpha_\mathrm{IMF}$ and N$_\mathrm{Ia}$, and 0.6 for x$_\mathrm{out}$; it is $\sim 0.8$ for SFE, SFR$_\mathrm{peak}$ and f$_\mathrm{corona}$, implying that the observational constraints place only weak constraints on those parameters. Only for $\tau_\mathrm{Ia}$ there is no constraining signal, because of Sun+ alone having limited information on the delay time of SN\,Ia. We also see that the SFR$_\mathrm{peak}$ (and to a lesser degree SFE and f$_\mathrm{corona}$) has a distorted distribution. The reason is that the SFR is a highly non-linear function of our SFR$_\mathrm{peak}$ parameter by which it is parametrized. The effective change of the SFR is much stronger for lower values of SFR$_\mathrm{peak}$ which is why the inferred parameter distribution spreads out to higher values (cf. Figure\,\ref{fig:SFR}).

For the alternative yield set (red diamonds) the inference is fairly consistent with small biases for N$_\mathrm{Ia}$ ($-1.2\sigma_\mathrm{prior}$), SFE ($+1.5\sigma_\mathrm{prior}$) and SFR$_\mathrm{peak}$ ($-1\sigma_\mathrm{prior}$). The uncertainties of the inferred marginalized posterior distribution is very similar except for the SFR$_\mathrm{peak}$, the reason being the non-linearity of this parameter. When looking at the feedback of the two yield sets in Figure\,\ref{fig:yield} we can qualitatively explain the behaviour of inferred N$_\mathrm{Ia}$. Since the \citet{Chieffi2004} CC-SN yields produce more Fe, \CP~decreases the number of SN\,Ia in order to match the $\alpha$/Fe abundances. In practice it is much more complicated to establish a causal connection between model assumptions (like the yield) and inferred parameters. It could be, that the decreased N$_\mathrm{Ia}$ is just a consequence of the decreased SFR$_\mathrm{peak}$, as both parameters are positively correlated (see Figure\,\ref{fig:parameter_space}).

The best (worst) maximum posterior values from 10 MCMC inferences are -5.5 (-13.7) for the default yield set, and significantly worse, -34.4 (-70.4), when using the alternative yield set: the mock data clearly prefer the default yield set, from which they were synthesized.
\subsection{\CP\  parameters constrained by Sun+}
\label{ch:separate_zones}
\begin{figure*}
\resizebox{\hsize}{!}{\includegraphics{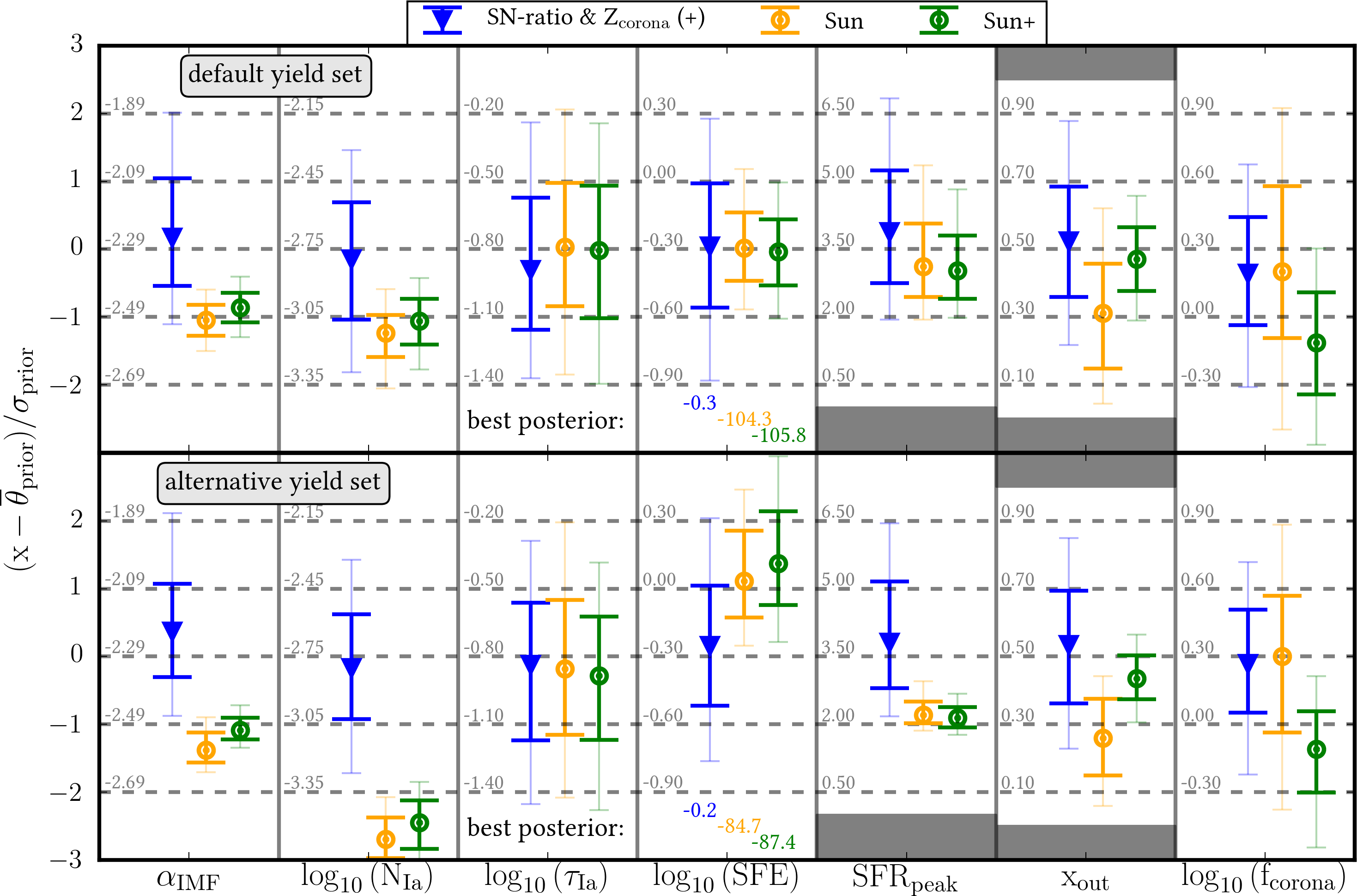}}
\caption{As in Figure\,\ref{fig:mock_default} we show the \CP~parameter constraints that can be derived from simple observational constraints, but now using "real" data: a) the SN ratio in external galaxies and Z$_\mathrm{corona}$ (blue, '+'); the Sun's abundances alone (orange), and Sun+ (green). The top/bottom row shows the results for the default/alternative yield set, as listed in Table\,\ref{tab:results_small}.
}
\label{fig:multiple_obs}
\end{figure*}

We now proceed to apply \CP\ to different subsets of the actual data. 
In Figure\,\ref{fig:multiple_obs} the inferred parameter distributions are shown when using the following subsets of our observational constraints: SN-ratio together with corona metallicity (4, 5 of Table\,\ref{tab:obs_constraints}, blue), Sun-only (1, in orange) and Sun+ (1,4,5, in green). The upper (lower) panel shows the results for the default (alternative) yield set.

For Sun+ with the default yield set the inferred median parameter values are compatible with $\overline{\vec{\theta}}_\mathrm{prior}$ except for $\alpha_\mathrm{IMF}$, N$_\mathrm{Ia}$ and $\mathrm{f}_\mathrm{corona}$ where it is $\approx\overline{\vec{\theta}}_\mathrm{prior}-\sigma_\mathrm{prior}$. The achieved precision is remarkable with the number of CC-SN (SN\,Ia) per 1000\,M$_\odot$ being constrained to $6.6\pm0.7$ ($0.9\pm0.2$) and similarly the mass fraction of the IMF that is turned into CC-SNe ($m>8\,\mathrm{M}_\odot$) being constrained to $10.3\,\%\pm1.2\,\%$. Note that both yield sets obtain comparable results (modulo the offset already seen in the synthetic data test, cf. Figure\,\ref{fig:mock_default})

If we only use SN-ratio and corona metallicity as constraints (black triangles), we see that the posterior distribution of the parameters is not departing much from the prior distribution, $\overline{\vec{\theta}}_\mathrm{prior}$. A small shift is visible for the corona mass normalisation, $\mathrm{f}_\mathrm{corona}$, which influences the corona metallicity and for the high-mass slope of the IMF, $\alpha_\mathrm{IMF}$, and the SFR$_\mathrm{peak}$ which are influencing the SN-ratio. The effect of SN-ratio and corona metallicity is small because only two data points (with large uncertainties) are "competing" with 7 parameter priors. Still when adding these additional constraints to the Sun (i.e. comparing Sun to Sun+), we see that for both yield sets the precision of x$_\mathrm{out}$ and f$_\mathrm{corona}$ increases. At the same time the former increases and the latter decreases by $1\sigma_\mathrm{prior}$.

With respect to our sub-title it is in particular remarkable that the Solar elemental abundances (i.e. the orange marker in Figure\,\ref{fig:multiple_obs}) alone, put very tight constraints on the high-mass slope of the IMF and the incidence of SN\,Ia. Meaning that within our \CP~framework the posterior distributions of these parameters are much narrower than their respective priors. However this comes with the caveat that we have to trust our assumptions especially the applied yield sets which visibly bias our results (cf. the two panels of Figure\,\ref{fig:multiple_obs}).

\begin{figure}
\resizebox{\hsize}{!}{\includegraphics{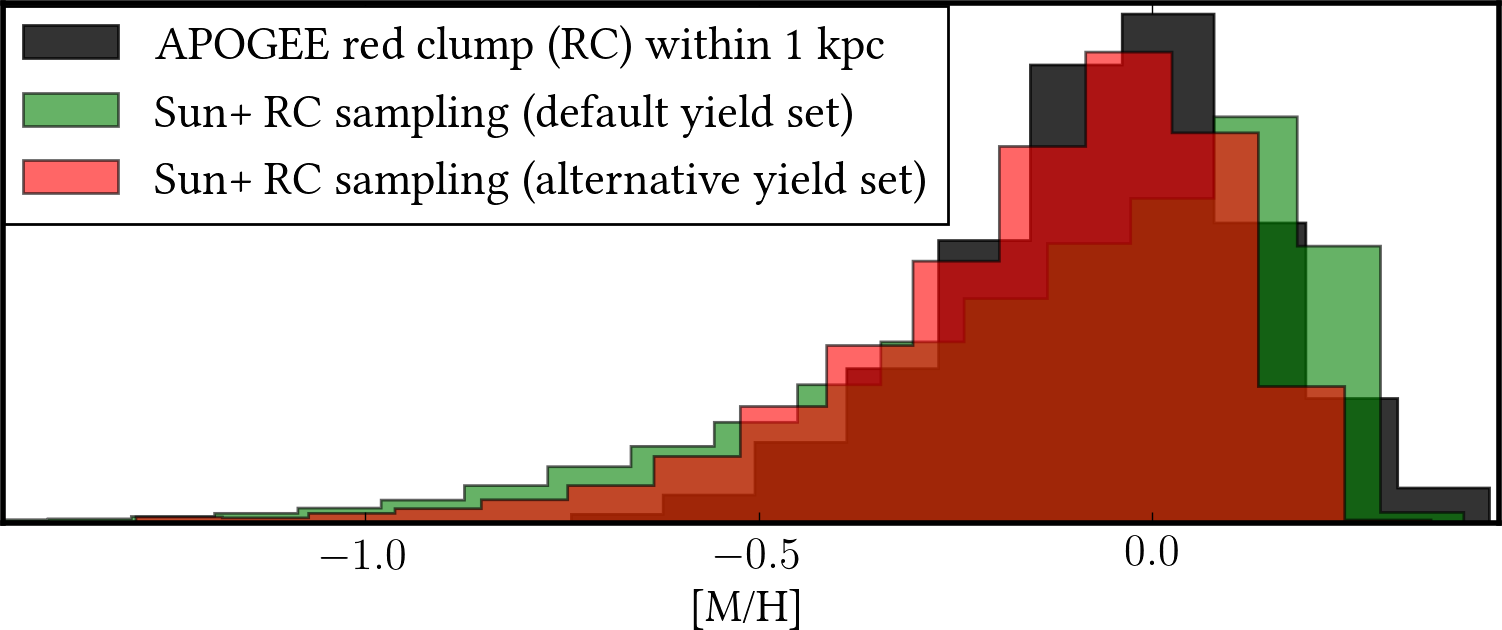}}
\caption{Metallicity distribution function (MDF) of APOGEE RC \citep{Bovy2014b} stars (black) within 1\,kpc of the Sun compared to \CP~predictions for the default (green) and alternative yield set (red) with Sun+ as observational constraint. The \CP~ISM abundances were weighted with the corresponding SFR and the RC age distribution to obtain synthetic observations. Note that the APOGEE MDF was not used during the fitting procedure.}
\label{fig:mdf}
\end{figure}

In Figure\,\ref{fig:mdf} we compare the histograms of predicted MDF of the default (in green) and alternative yield set (in red) with APOGEE red clump (RC) stars \citep{Bovy2014b} from within 1\,kpc of the Sun (in black). In order to produce synthetic observations we weight Chempy ISM abundances with the associated SFR and the age distribution of RC stars, as described in \citet{Just2015a}. Both predicted distributions qualitatively match the local APOGEE RC sample which peaks around Solar metallicity. The default yield set peaks slightly higher and both predicted MDFs peter out later at about -1\,dex whereas the observations only reach to about -0.6\,dex. As mentioned in Section\,\ref{ch:MDF} we do not believe that there exists a one-to-one mapping between the \CP~Solar zone and the Solar neighbourhood. E.g. the low metallicity tail containing old stars could be lost to the present-day Solar neighbourhood due to dynamical processes.
\subsubsection{Parameter correlations}
\label{ch:correlation}
\begin{figure*}
\resizebox{\hsize}{!}{\includegraphics{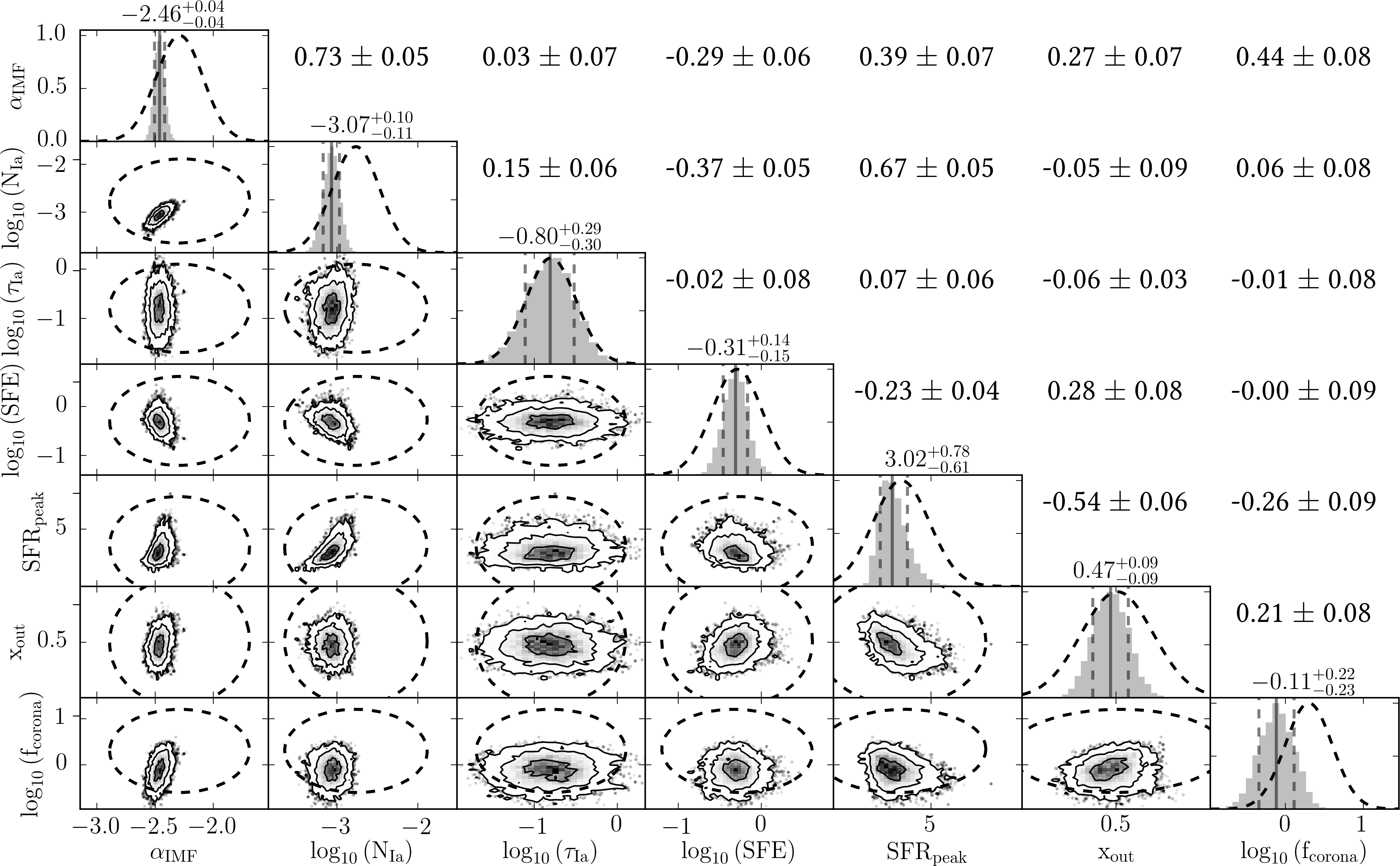}}
\caption{Marginalized parameter distribution derived from the Sun+ constraints, in comparison to the prior distribution. Each contour plot in the lower left shows the projected 2D parameter density distribution with 1,2 and 3$\sigma$ contours, with individual PDF sample points beyond \citep{Foreman-Mackey2016}. The 3$\sigma$ ellipse from the prior run is shown in dashed black. The respective correlation coefficients (Pearson's r) are given at the upper right together with the standard deviation from 10 identical inferences. Histograms of the marginalized parameter distributions are given on the diagonal together with the Gaussian distribution of the prior which is shown in dashed black. The median and the 16 \& 84 percentiles of each parameter PDF are given above the histogram and indicated as solid and dashed grey lines.}
\label{fig:parameter_space}
\end{figure*}

In Figure\,\ref{fig:parameter_space} the corner plot for the inferred parameter distribution is shown when using \CP~with the default yield set and Sun+ as observational constraint. This allows us to investigate the mechanics of \CP~via parameter correlations.

The strongest correlation ($0.73\pm0.05$) is between $\alpha_\mathrm{IMF}$ and $\mathrm{N}_\mathrm{Ia}$ with more CC-SNe also demanding for more SNe\,Ia in order to get the abundance plateaus of $\alpha$-elements and iron-peak elements right. Both parameters are also positively correlated with $\mathrm{SFR}_\mathrm{peak}$ because a later peak of SFR means that more material is turned into low metallicity stars which themselves produce less metals. Likewise, the estimates of the outflow fraction and of the corona mass are anti-correlated with the SFR$_\mathrm{peak}$ (-0.54, -0.26), because a larger outflow can be compensated by earlier enrichment.

For $\mathrm{x}_\mathrm{out}$ and $\mathrm{f}_\mathrm{corona}$ only $\alpha_\mathrm{IMF}$ has a positive correlation (0.27, 0.44) so that more produced metals can be compensated by more outflow and a larger corona mass to be mixed with, in order to satisfy the Solar abundance and the corona metallicity data. $\alpha_\mathrm{IMF}$ and $\mathrm{N}_\mathrm{Ia}$ have a small negative correlation with SFE (-0.29, -0.37) due to more ISM gas will need more feedback for the same enrichment. Similarly the SFE is negatively correlated with SFR$_\mathrm{peak}$ (-0.23) and positively correlated with x$_\mathrm{out}$ (0.28).

By rerunning our inference 10\,times we can give standard deviations for the correlation coefficients and at the same time we find that the inferred median parameter values, $\overline{\theta}_\mathrm{posterior}$, are very stable (to within $0.1\sigma_\mathrm{posterior}$) most likely due to parameter correlations.

\subsubsection{Inferred Element Production by the Different Nucleosynthetic Processes}
\begin{figure*}
\resizebox{\hsize}{!}{\includegraphics{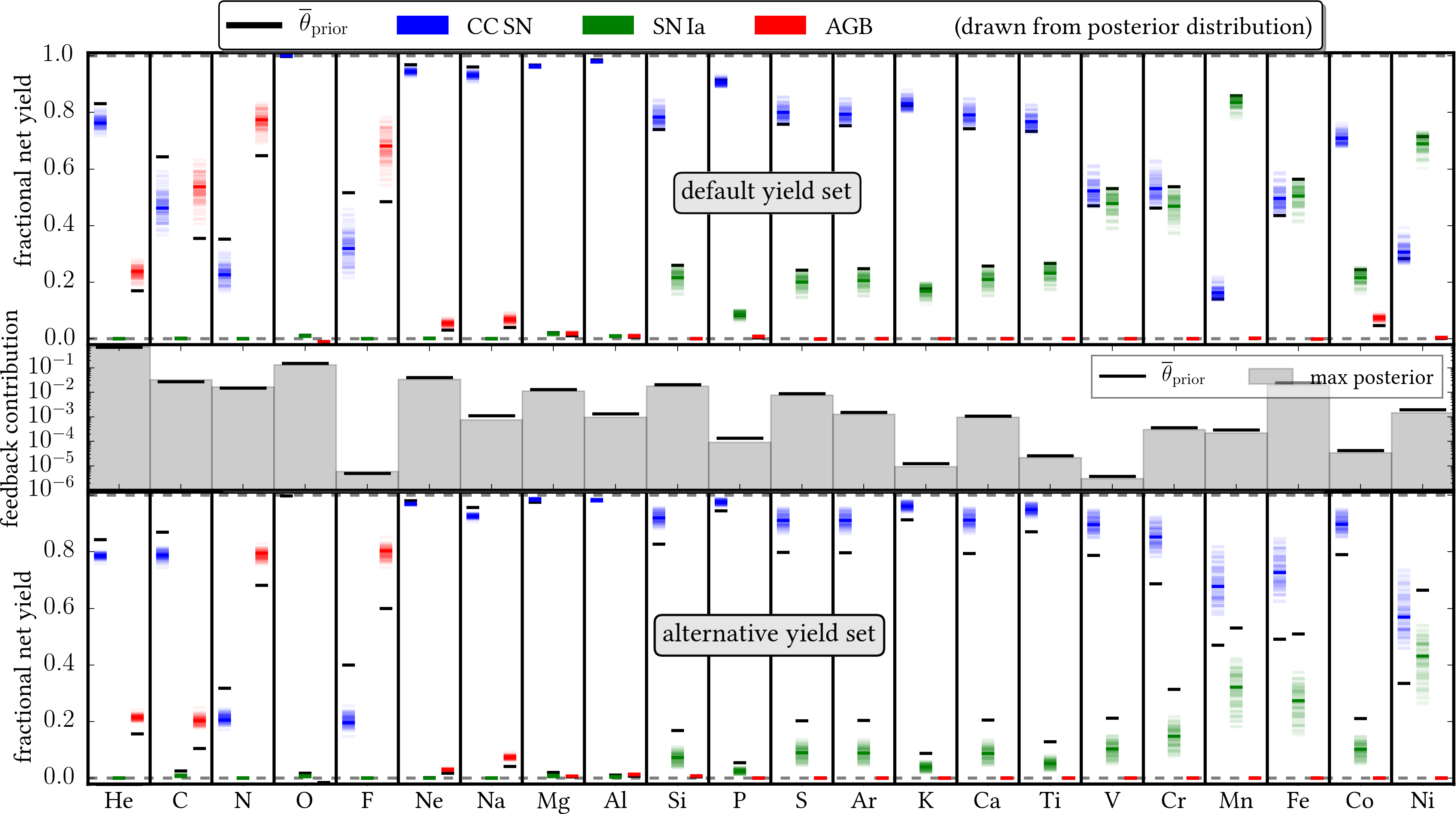}}
\caption{This Figure illustrates which fraction of light elements are produced by which nucleosynthetic channel, when the Chempy parameters are constrained by the Sun+
observations. The fractional yield of CC-SN, SN~Ia and AGB is shown in blue, green and red, respectively (in solid for the maximum posterior parameters, and transparently for 100 results from the converged MCMC). The upper (lower) panel shows the result using the default (alternative) yield set. The bars in the middle panel indicate each element's contribution to the total net yield (for the maximum posterior of the default yield set, for comparison, the prior feedback contribution is given in black lines but the difference is hardly visible on the log scale). The results from $\overline{\vec{\theta}}_\mathrm{prior}$ parameters are shown as black lines. These are best suited when looking for differences in the yield sets because the colored lines coming from $\theta_\mathrm{posterior}$ are superimposed with the effect of having different \CP~parameters. It is important to note that the CC-SN yield of the alternative yield set \citep{Chieffi2004} are gross, not net yields. Therefore the feedback includes unprocessed Solar scaled material.}
\label{fig:sun_default_feedback_elements}
\end{figure*}

In order to see the contribution from each nucleosynthetic channel to each element we have plotted their fractions in Figure\,\ref{fig:sun_default_feedback_elements} for the default (upper panel) and the alternative (lower panel) yield set (both optimized for Sun+). In the middle panel the mass fraction of each element's contribution to the total feedback is shown for the default yield set. Contributions from CC-SNe, SNe\,Ia and AGB stars are given in blue, green, red, respectively. There are 100 transparent lines from the posterior distribution that are plotted, with the most probable posterior fractions shown in the solid line. The fractions when using $\overline{\vec{\theta}}_\mathrm{prior}$ parameter values are given in black (they should be used to see differences between the two yield sets as the colored lines in the upper and lower panel have parameter changes superimposed onto the change in yield set). Because we marginalize out \CP~parameters we can see a more realistic range of possible fractional element production per nucleosynthetic process than any other study before. Cf. \citet[fig.~10]{Andrews2017} for a nucleosynthetic contribution per element over metallicity, albeit for a single set of (hyper-)parameters. 

From Figure\,\ref{fig:sun_default_feedback_elements} we learn that AGB stars mainly contribute to He, C, N and F enrichment (if our yield tables are applicable and if we neglect the s-process elements). The contribution to Carbon varies strongly between the two yield sets, because of the very high C production from the \citet{Chieffi2004} CC-SNe yields (cf. Figure\,\ref{fig:yield}), but is never as low as found in \citet[fig.~5]{Henry2000}, who uses \citet{Maeder1992} CC-SN and \citet{VandenHoek1997} AGB yields. Similarly their 90\,\% N contribution from AGB stars is a bit higher than our 75-80\,\% range. Our \CP\ modeling also provides theoretical evidence, that the main source of cosmic F are AGB stars \citep[c.f.][]{Recio-Blanco2012,Jonsson2014,Pilachowski2015}, with minor contributions from CC-SN.

For CC-SNe we see that only O and Mg are the "most pure" $\alpha$-elements, in the sense that their feedback is almost exclusively coming from CC-SNe. We see that Si, S, Ca and Ti have non-negligible contribution from SNe\,Ia. Of interest is also the fractional contribution of iron-peak elements from SNe\,Ia. Their Fe contribution ranges from 30 to 50\,\% as in \citet{Timmes1995}. The difference is mainly due to lower SN\,Ia normalisation of the alternative yield set, though the difference between the more physically motivated \citet{Seitenzahl2013} vs. the older \citet{Thielemann2003} SN\,Ia yields is also strong. We also see that Mn or Ni are better indicators for SN\,Ia incidence than Fe.

Overall we find a large uncertainty for each element's origin: especially for C, V, Cr, Mn and Ni the fractional contribution differs strongly between the different yield sets. Figure\,\ref{fig:yield} together with Figure\,\ref{fig:sun_elements} show the potential diagnostic power of \CP~in confronting nucleosynthetic yields and chemical evolution models with observations (cf. \citet{Mikolaitis2016}). 
\subsubsection{\CP\ constraints from B-stars+ and Arcturus+}
In Figure\,\ref{fig:multi-summary} the inferred parameter distributions are shown for B-stars+ (\{2,4,5\} of Table\,\ref{tab:obs_constraints}, in blue) and Arcturus+ (\{3,4,5\}, in red) and can be compared to the Sun+ constraints in green. Overall the inferred parameters for default and alternative yield set for B-stars+ and Arcturus+ are comparable and mostly within 1\,$\sigma_\mathrm{prior}$. Exceptions are the lower SN\,Ia normalisation for Arcturus+ with the default yield set and the longer SN\,Ia delay for Arcturus+ with the alternative yield set illustrating how \CP~is "struggling" to fit an $\alpha$-enhanced star within its physical model.

Even though the B-star and Arcturus constraints involve fewer element abundances (i.e. fewer constraining data points), the precision of inferred parameters is comparable to the one from Solar abundances, indicating that redundant information is contained in different elements.

The inferred ISM parameters of all three cases show an interesting trend for both yield sets. The SFE is lowest with Arcturus+, higher for B-star+ and highest for the Sun+. The $\mathrm{SFR}_\mathrm{peak}$ is earlier for Sun+ and later for B-stars+ (Arcturus+ being undecided). The outflow fraction is high for B-star+ and Arcturus+ but low for Sun+. And also the corona normalisation is similar for B-stars+ and Arcturus+ and only half as big for Sun+. These differences may be affected by the cross-correlation of parameters (the reaction of the ISM parameters to having different SSP parameters, cf. Figure\,\ref{fig:parameter_space}, but it would also mesh with the commonly invoked narrative:
\begin{itemize}
\item The Sun originated further inside the Galaxy (inner thin disk), where the SFE and the gravitational potential was higher and stars formed earlier than at its present day position.
\item B-stars (i.e. the local ISM) originate at a larger radius, with a later peak in SFR,
less retention of feedback material in the ISM, and an overall lower SFE (outer thin disk).
\item Arcturus originated in an environment with many CC-SNe and few SNe\,Ia, a large outflow and a lowered SFE (thick disk).
\end{itemize}

\subsubsection{How well are stellar abundances reproduced?}
\label{ch:solar_abundances}
\begin{figure*}
\begin{center}
\includegraphics[width=\textwidth]{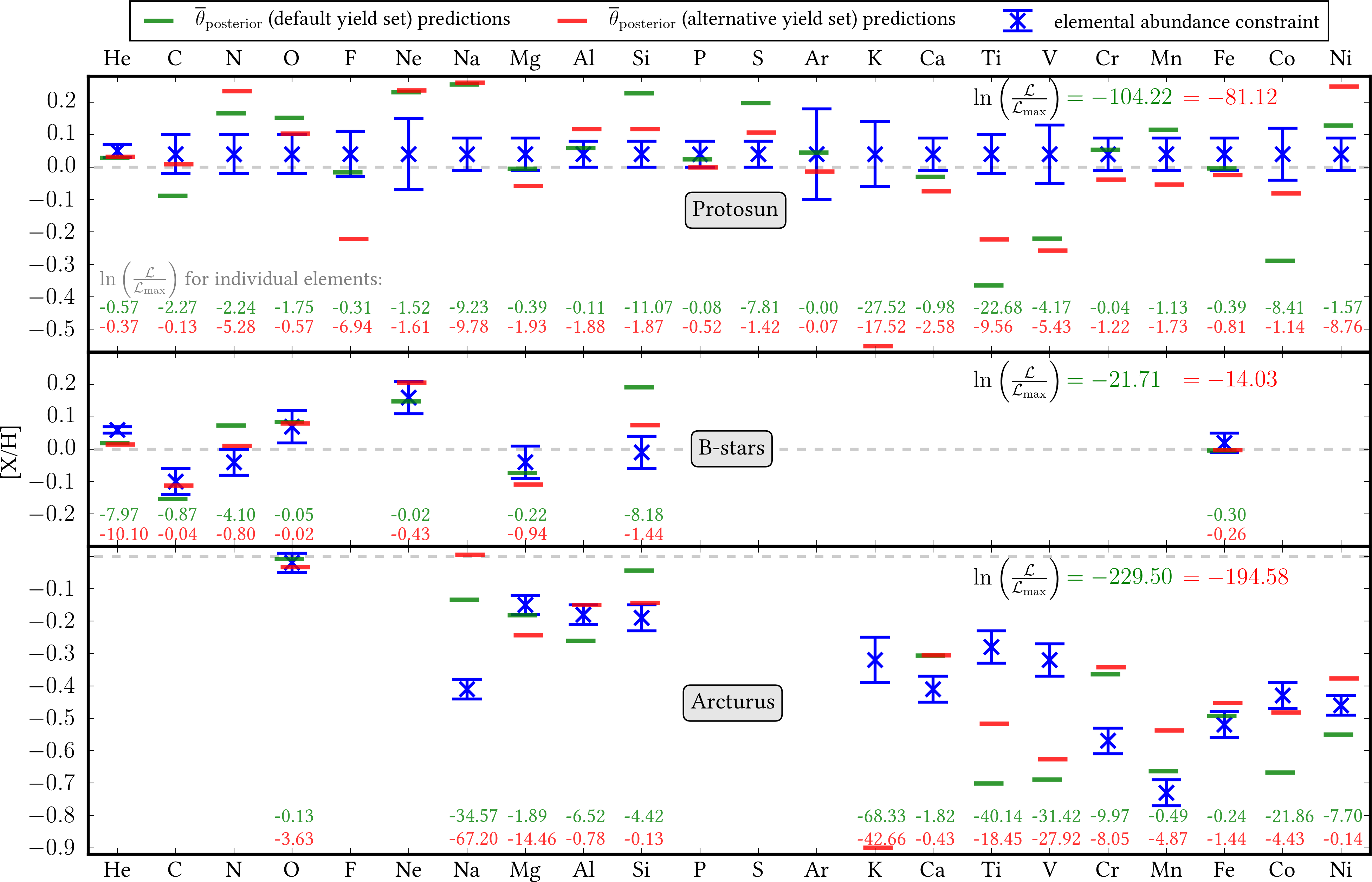}
\caption{\CP~abundance predictions (for the medians of the marginalized posterior PDFs, $\overline{\theta}_\mathrm{posterior}$), compared to the observed stellar abundances. The top row compares the predicted (red, green) and observed (blue) abundances for the Sun, after optimizing for Sun+ constraints; the central panel shows an analogous comparison for the B-stars, and the bottom panel for Arcturus. Predictions for the default yield set are shown in green; for the alternate yield set in red. The log likelihood coming from each individual element as well as the sum of all are indicated. In general, the default and alternate yield sets fit comparably well (or poorly). The predictions for some elements, e.g. Na, K Ti and V are poor, for both yield sets and for Sun+ and Arcturus+. This implies that some abundances simply cannot be reproduced in the context of GCE models, even if a model as flexible as Chempy is applied.} 
\label{fig:sun_elements}
\end{center}
\end{figure*}

In Figure\,\ref{fig:sun_elements} we want to investigate how well \CP\  predictions do reproduce the observational constraints. We show the stellar abundance data and predictions coming from the median marginalized parameter values of the posterior PDF, $\overline{\theta}_\mathrm{posterior}$, when optimizing for the respective stellar abundance and the additional observational constraints (i.e. Sun+, B-stars+, Arcturus+) with the default (green lines) and the alternative (red lines) yield set.

Figure\,\ref{fig:sun_elements} shows that many element abundances can be reproduced well, by both yield sets. However, some abundances  cannot be explained (within a  factor of 2) by either the default, or the alternate yield set, 
even with the fitting flexibility that \CP\ otherwise affords. 
Overall, the maximum obtained posterior is always somewhat better when using the alternative yields, implying that the abundances (together with the SN-ratio and the corona metallicity) are better reproduced. We attribute this to the Solar scaled feedback of unprocessed material from \citet{Chieffi2004} CC-SNe yields whereas for \citet{Nomoto2013} we use net yields (i.e. the unprocessed feedback is composed of the stellar birth material), though the selection of elements can also change the posterior ranking of the yield tables. Remarkably the best posterior of Arcturus+ is lower than from Sun+ with both yield sets, even though the Arcturus constraints encompass fewer elements. This may be attributable to
the formally higher precision of the Arcturus data. But it may 
also imply that \CP's implementation lacks the ability to produce $\alpha$-enhanced abundance patterns at such late times. Other authors circumvent this by e.g. invoking metallicity dependent SN\,Ia rates \citep{Kobayashi2006}.

If we perturbed the protosolar abundances by their observational error and calculate the likelihood with itself $10^5$\,times the median and 16\,\&\,84 percentiles of the log likelihood would be
\begin{equation}
\ln\left(\frac{\mathcal{L}}{\mathcal{L}_\mathrm{max}}\right) = -10.7^{+2.9}_{-3.6}.
\end{equation}  
Since our log likelihood with the default yield set is -104.2 we are far away from reproducing the protosolar abundances precisely (or accurately), even though we marginalize out all \CP\  parameters. The range of possible reasons: shortcomings of the yield tables \citep{Rauscher2016}, inaccurately determined abundances \citep{Bergemann2012}, inhomogeneous mixing of the ISM metals \citep{Venn2012} or nucleosynthetic channels which we have not included, e.g. sub-luminous SN\,Ia \citep{Pakmor2010}, sub-Chandrasekhar SN\,Ia \citep{Woosley1994}.
Additionally \CP~ is a very simple model (with its one-zone) and most of the parameters are assumed to be constant over time (x$_\mathrm{out}$, $\alpha_\mathrm{IMF}$, N$_\mathrm{Ia}$, SFE). These model assumptions may not be good approximations for the whole Milky Way evolution. Also the functional form of the SFR, the IMF and the SN\,Ia DTD are quite restrictive. Nevertheless, we can infer problems with the yield sets from consistent inability of \CP\  to reproduce certain elemental abundances which we will discuss next.

Optimizing the parameters for Sun+ with the default yield set and using the predictions from the median posterior parameter values, the following elements are more than 3$\sigma_\mathrm{obs}\,(\hat{=}\ln(\mathcal{L}/\mathcal{L}_\mathrm{max})=-4.5)$ away from the observations and will be given in units of $\sigma_\mathrm{obs}$: K\,($-7.4$), Ti\,($-6.7$), Si\,($+4.7$), Na\,($+4.3$),  Co\,($-4.1$) and S\,($+4.0$). Looking at Figure\,\ref{fig:sun_default_feedback_elements} we see that all of those elements are at least produced to 2/3 by CC-SN and all of those except Na have a 1/5 contribution from SN\,Ia.

For the alternative yield set the obtained likelihood is somewhat better ($-81.1$ compared to $-104.2$). Here the elements with high deviations are: K\,($-5.9$), Na\,($+4.4$), Ti\,($-4.4$), Ni\,($+4.2$), F\,($-3.7$), V\,($-3.3$) and N\,($-3.2$). The contributions to those elements from CC-SNe are significantly higher (except for F) partly due to smaller SN\,Ia normalisation and also due to \cite{Chieffi2004} only being implemented as gross yields in \CP~so far, including Solar scaled feedback.

We note that \CP~slightly over-predicts O, Ne for the Sun. These elements are part of the Solar abundance problem \citep{Serenelli2016} and our adopted values \citep{Asplund2009} are at the lower limit of the debated abundance range \citep{Caffau2011}, meaning that an increase could remedy our over-prediction. Vice versa \CP~could potentially identify offsets in abundance determination, provided the yield sets being accurate.

In the middle panel of Figure\,\ref{fig:sun_elements} the same is plotted for B-stars+ (using a smaller set of elements). Again the log likelihood of the default yield set is worse, with  Si ($+4.0$) and He ($-4.0$) showing the largest deviations. For the alternative yield set only He is off ($-4.5$) more than $3\,\sigma$ and otherwise the predictions would be consistent with the B-stars abundances. Interestingly, the He predictions are a bit low for the Sun and strongly under-abundant in B-stars for both yield sets. Again this could be remedied by an increase in the Solar He abundance, which would also be more consistent with results from helioseismology \citep{Basu2004}.

Arcturus+ results are plotted in the lower panel of Figure\,\ref{fig:sun_elements} (beware that [X/Fe] was used in the likelihood calculation, though [X/H] is depicted here). The alternative yield set gives better results and the deviation from the observations for both yield sets are stronger than for the Sun+ and B-stars+. More than half of the elements are $3\,\sigma$ outliers for both yield sets. For the default yield set the largest deviations are coming from K ($-11.7$), Ti ($-9.0$), Na ($+8.3$), V ($-7.9$), Co ($-6.6$), Cr ($+4.5$), Ni ($-3.9$) and Al ($-3.6$). For the alternative yield set the outliers are Na ($+11.6$), K ($-9.2$), V ($-7.5$), Ti ($-6.1$), Mg ($-5.4$), Cr ($+4.0$) and Mn ($+3.1$).

Since uncertainties of the abundances are expected to be much lower, we attribute the consistent (for both yield sets) under-prediction of K, Ti, V and over-prediction of Na to the yield tables. Alternatively it could mean that we are missing a non-negligible nucleosynthetic channel \citep{Mernier2016}. \citet{Battistini2015} found for V that it should behave as an SN\,II-like element, so that the deficiency could be attributed to the \citet{Nomoto2013} CC-SNe yields. As for K, \citet{Ventura2012} speculate that super-AGB stars could be an important source. From Figure\,\ref{fig:yield} we see that neither of our yield sets has K contribution from AGB stars.

 The under abundance of K, Ti and V is also found in other chemical evolution studies \citep{Goswami2000,Henry2010,Kobayashi2006,Andrews2017}. Similarly, \citet{Francois2004} found an under abundance of Ti using the \citet{Woosley1995} and \citet{Iwamoto1999} yields, though K works for them. Looking at \citet[Fig.\,29]{Sukhbold2015} K, Ti, V seem to fit in newer CC-SNe models as well as Na, albeit only for Solar metallicity yields.

The maximum of the posterior PDF can be used as an indicator which yield set best reproduces observations. This could help to discriminate between different nucleosynthetic feedback models and could also be used to infer empirical yields sets.
\subsection{Multi-zone scheme}
\label{ch:multizone}

The advent of big spectroscopic surveys, eventually providing the abundances for millions of stars,
in principle hold the key to constrain all parameters involved in the chemical enrichment. Yet matching the
abundances of stars across the Milky Way with a single one-zone model, must be a poor model approximation.
This limitation already manifests itself in the MCMC runs where we put all three stars in the same zone (mutual single-zone run, cyan in Figure\,\ref{fig:multi-summary}). We see that the best achieved posteriors for both yield sets are much worse than if we just added the posteriors of the individual runs. The retrieved parameters depart strongly from $\overline{\vec{\theta}}_\mathrm{prior}$ illustrating the "tension" arising from the assumption that Sun's and Arcturus's abundance patterns were produced in the same chemical enrichment zone. Also for the default yield set the MCMC only finds a 'pathological' parameter configuration, where the SFR ceases after $\sim$8\,Gyr and the $\alpha$-enhancement decreases due to a few remaining SNe\,Ia (see solid cyan line Figure\,\ref{fig:apogee}).

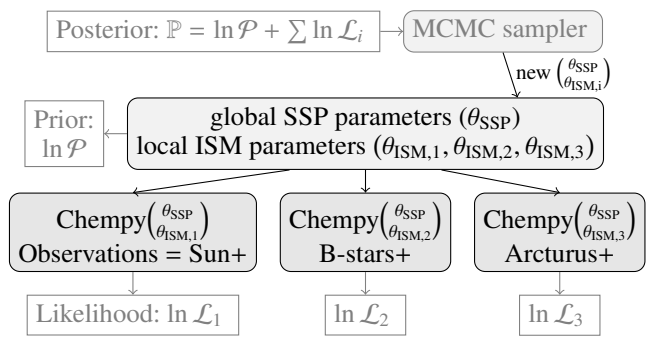
\begin{figure}
\begin{center}
\begin{tikzpicture}[scale=2]
 \node[align = center] (SSP) [rectangle,
            rounded corners, draw, fill = black!5] 
 {global SSP parameters $\left(\vec{\theta}_\mathrm{SSP}\right)$\\
 local ISM parameters $\left(\vec{\theta}_\mathrm{ISM,1},\vec{\theta}_\mathrm{ISM,2},\vec{\theta}_\mathrm{ISM,3}\right)$
};
 \node[align = center, below = 0.3 of SSP] (CAS) [rectangle,draw=black,
            rounded corners, fill=black!10] 
 {Chempy${\theta_\mathrm{SSP}\choose\theta_\mathrm{ISM,2}}$ \\
 B-stars+
};
 \node[align = center, left = 0.3 of CAS] (SUN) [rectangle,draw=black,
            rounded corners, fill=black!10] 
{Chempy${\theta_\mathrm{SSP}\choose\theta_\mathrm{ISM,1}}$ \\
 Observations = Sun+
};
 \node[align = center, right = 0.3 of CAS] (ARC) [rectangle,draw=black,
            rounded corners, fill=black!10] 
{Chempy${\theta_\mathrm{SSP}\choose\theta_\mathrm{ISM,3}}$ \\
 Arcturus+
};
 \node[align = center, below = 0.3 of SUN] (LSUN) [rectangle,
             draw, gray] 
 {Likelihood: $\ln\mathcal{L}_1$
};

 \node[align = center, below = 0.3 of CAS] (LCAS) [rectangle, draw,
             gray] 
 {$\ln\mathcal{L}_2$
};
 \node[align = center, below = 0.3 of ARC] (LARC) [rectangle,
             draw, gray] 
 {$\ln\mathcal{L}_3$
};
 \node[align = center, left = 0.3 of SSP,xshift=0cm] (PRIOR) [rectangle,
             draw, gray] 
 {Prior:\\
  $\ln\mathcal{P}$
};

  \node[align = center, above = 0.6 of SSP, xshift = -2.0cm] (POSTERIOR) [rectangle,
              draw, gray] 
  {Posterior: $\mathbb{P}=\ln\mathcal{P} + \sum\ln\mathcal{L}_i$
 }; 
   \node[align = center, right = 0.3 of POSTERIOR] (MCMC) [rectangle,
              rounded corners, draw, gray, fill=black!5] 
   {MCMC sampler
  }; 
 \path ([xshift = 0cm]SSP.west)     	edge[bend left=0, ->,gray] node[gray,yshift=0.2cm] {} (PRIOR.east);
 
  \path (POSTERIOR.east)     	edge[bend left=0, ->,gray] node[gray,yshift=0.2cm] {} (MCMC.west);
  \path ([xshift = -0.0cm]MCMC.south)     	edge[bend left=0, ->] node[yshift=0.0cm,xshift=0.7cm] {\tiny new ${\vec{\theta}_\mathrm{SSP}\choose\vec{\theta}_\mathrm{ISM,i}}$} ([xshift = 1cm]SSP.north);  
 \path (SUN.south)     	edge[bend left=0, ->,gray] node[gray,yshift=0.2cm] {} (LSUN.north);
  \path (CAS.south)     	edge[bend left=0, ->,gray] node[gray,yshift=0.2cm] {} (LCAS.north);
   \path (ARC.south)     	edge[bend left=0, ->,gray] node[gray,yshift=0.2cm] {} (LARC.north);
    \path ([xshift = -0.5cm]SSP.south)     	edge[bend left=0, ->] node[gray,yshift=0.2cm] {} ([xshift = 0cm]SUN.north);  
    \path ([xshift = 0.cm]SSP.south)     	edge[bend left=0, ->] node[gray,yshift=0.2cm] {} ([xshift = 0cm]CAS.north);  
       \path ([xshift = 0.5cm]SSP.south)     	edge[bend left=0, ->] node[gray,yshift=0.2cm] {} ([xshift = 0cm]ARC.north);

\end{tikzpicture}
\caption{multi-zone scheme to use constraints from multiple stars.}
\label{fig:multizone_scheme}
\end{center}
\end{figure}

We explore one obvious step to overcome this unrealistic assumption by generalizing
\CP\ to a {\it multi-zone} model, as depicted in Figure\,\ref{fig:multizone_scheme}. We run three \CP~models simultaneously, one for each star, where we require that all three zones share the same SSP parameters (i.e. share the same global stellar physics, cf. Table\,\ref{tab:abbreviations}) but each zone can have their individual ISM parameters (i.e. their own local ISM history). Then we add up their log likelihoods and sample their common posterior PDF over the increased parameter space ($2\times\mathrm{N}_{\vec{\theta}_\mathrm{ISM}}=8$ additional parameters)\footnote{For simplicity we treat the 15\,parameters of the multi-zone scheme equally, though for an increased number of stars treating them hierarchically (e.g. in a Gibbs sampler) separating global from local parameters will prove advantageous.}.

\begin{figure*}
\begin{center}

\includegraphics[width=\textwidth]{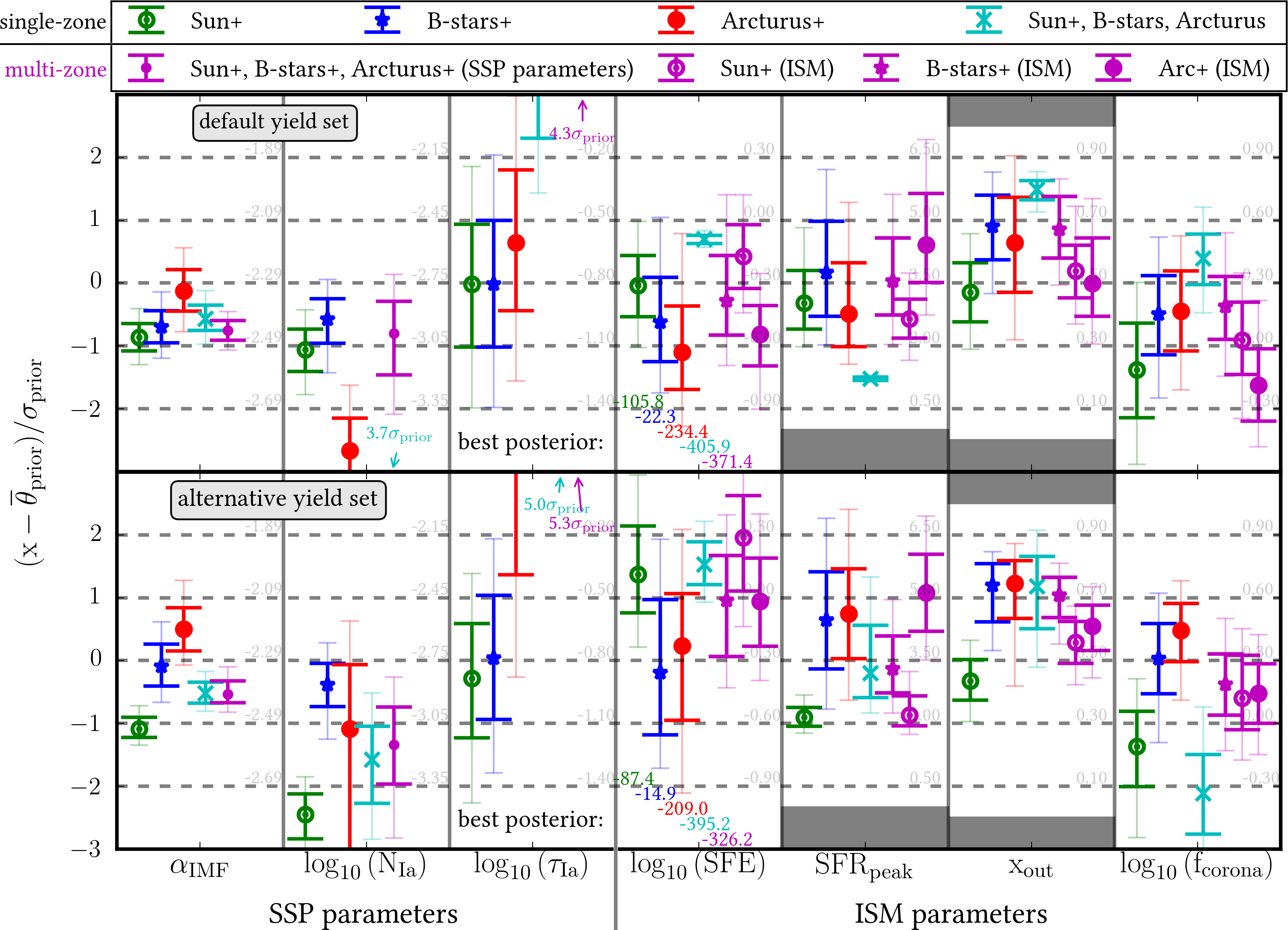}
\caption{This figure compares the inferred model parameters (PDFs) when fitting the three sets of observational constraints separately (Sun+, B-star+, Arcturus+), to the case when these constraints are fit simultaneously. For fitting the constraints simultaneously, we consider two regimes: first fitting a single-zone model, i.e. a single set of SSP and ISM parameters (cyan); or fitting a multi-zone model, where there is a single set of stellar physics parameters, but Sun+, B-stars+ and Arcturus+ each (can) have their own ISM pre-history (magenta). For the SN\,Ia delay parameter some results are outside of the plotted $3\sigma_\mathrm{prior}$ range but their values are given and can also be checked in Table\,\ref{tab:results_small}.} 
\label{fig:multi-summary}
\end{center}

\end{figure*} 

In Figure\,\ref{fig:multi-summary} we show results from the multi-zone scheme in magenta using Sun+, Arcturus+ and B-stars+ simultaneously as observational constraint. We find that the (joint) SSP parameters $\alpha_\mathrm{IMF}$ and N$_\mathrm{Ia}$ are tightly constrained and consistent with the single-zone runs: we get $7.0\pm0.5$ ($7.7^{+0.7}_{-0.5}$) CC-SNe and $1.0\pm0.4$ ($0.7^{+0.4}_{-0.2}$) SNe\,Ia per 1000\,M$_\odot$ for the default (alternative) yield set; for our IMF that means $10.9^{+1.0}_{-0.9}\,\%$ ($12.2^{+1.5}_{-1.0}\,\%$) of the mass fraction will explode as CC-SNe for the default (alternative) yield set (cf. Figure\,\ref{fig:IMF}). 

The SN\,Ia time delay is departing from $\overline{\vec{\theta}}_\mathrm{prior}$ demanding an implausibly long delay of $3.1^{+0.4}_{-0.6}$\,Gyr ($6.0^{+0.8}_{-0.5}$\,Gyr) for the default (alternative) yield set. This shows that it is hard to reconcile the three abundance patterns within the physics and parametrization of \CP.
Either the SN\,Ia rate \citep{Kobayashi2006} or the IMF need to get metallicity dependent. Alternatively the delay time distribution could be improved by using a more realistic functional form \citep{Matteucci2001a} and more SN\,Ia channels \citep{Ruiter2009}. Though many more stars with different ages will be needed to reliably constrain the additional free parameters.

Remarkably, our multi-zone inference, based on either yield set has consistent, within $1\sigma_\mathrm{posterior}$, results for all parameters, except for $\tau_\mathrm{Ia}$ and Arcturus' $\mathrm{f}_{\mathrm{corona}}$. The resulting ISM parameters still support the idea of the three stellar abundances stemming from a different birth environment, i.e. inner thin disk, outer thin disk and thick disk (albeit for Arcturus+ x$_\mathrm{out}$ and f$_\mathrm{corona}$ the trend does no longer hold, probably due to correlations stemming from the changed SSP parameters).

In order to compare \CP\  predictions to observations in Figure\,\ref{fig:apogee} we are plotting the multi-zone (upper panels) and the single-zone (lower panels) predictions together with our constraining data in the [Mg/Fe] vs [Fe/H] plane (left panels). We show the same in the [Mg/Fe] vs time plane (right panels) and in both cases we add independent observation of $\sim$$20.000$ APOGEE giant stars \citep{Ness2016} to guide the eye.

\begin{figure*}
\resizebox{\hsize}{!}{\includegraphics{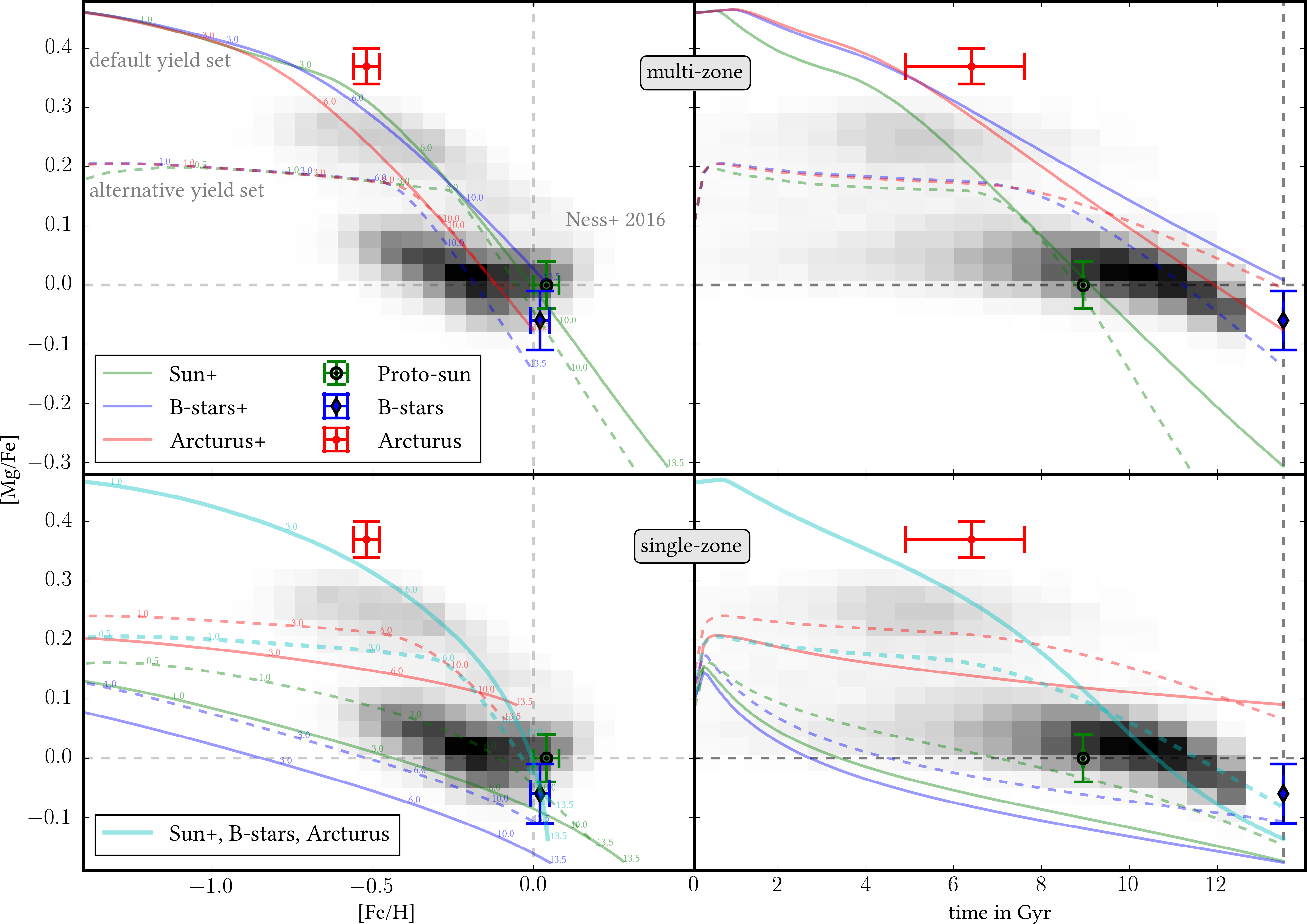}}
\caption{Tracks of the ISM abundances evolution implied by the \CP~fits, represented by a track in the [Mg/Fe] - [Fe/H] (left) and the [Mg/Fe] - time (right) panels. Shown are the predictions for the median inferred parameters $\overline{\theta}_\mathrm{posterior}$ coming from the multi-zone (upper panels) and the single-zones (lower panels, their colors are the same as in Figure\,\ref{fig:multi-summary}) inference. The default/alternative yield set are plotted in solid/dashed lines. The current simulation time is written in small letters next to the chemical abundance tracks. Our observational data, used for the fitting procedure, is shown with error bars. For comparison the distribution of $\sim$$20.000$ APOGEE giants from \citet{Ness2016} are shown as gray shaded area. The evolutionary tracks of the separate single-zones are flatter compared to the multi-zone tracks because of the almost 'prompt' SNe\,Ia. The evolutionary track of the solid cyan line, where Sun+, B-stars and Arcturus are included into a single zone using the default yield set, is so steep because the SFR ceases after $\sim$8\,Gyr, only leaving SNe\,Ia contributing to the enrichment. Keep in mind that all the elemental abundances (+SN-ratio \& Z$_\mathrm{corona}$) are fit for simultaneously, so a trade-off between all those constraints and the prior distribution is found by the MCMC.}
\label{fig:apogee}
\end{figure*}

We see that \CP\  is flexible enough to more or less fulfill the observational constraints, which actually happens in the time domain (right panels). Keep in mind that all the elemental abundances (+SN-ratio \& Z$_\mathrm{corona}$) are fit for simultaneously, so a trade-off between all those constraints and the prior distribution is found by the MCMC.

\CP's strong sensitivity to the SSP parameters (as evident from their small $\sigma_\mathrm{posterior}/\sigma_\mathrm{prior}$ compared to the ISM parameters) manifests in the similarity of the chemical enrichment patterns for the multi-zone approach. There the differences between the individual zones can to first order be explained by the different timing set by the (also quite sensitive) SFR$_\mathrm{peak}$ parameter. In both yield sets the plateau ends with the onset of the SNe\,Ia ($\approx$$3$\,Gyr for the default and $\approx$$6$\,Gyr for the alternative yield set), which are mostly prompt for the single-zone runs.

For the single-zone cases the evolutionary tracks look consistent with the behaviour of the APOGEE data. Even the cyan tracks, which incorporate Sun+, B-star and Arcturus, fit well, but the overall fit of this mutual single-zone to the other elements is much worse as apparent from the best posterior values. Naturally, the multi-zone case scores better when fitting all three abundance patterns simultaneously, albeit the Solar zone gets too metal-rich and too [Mg/Fe] poor. Additionally the track of the Solar zones take an alpha-enhanced route in the [Mg/Fe] vs [Fe/H] plane, though this picture changes in the [Mg/Fe] vs time plane where the Solar zones decrease $\alpha$-enhancement faster than the other zones. This points to the Solar zone being produced in an inner disk environment \citep{Wielen1996}.

Similarly, while we see an increase of Mg in the [Mg/Fe] vs. [Fe/H] plane for the APOGEE data, this does not hold when plotting in the [Mg/Fe] vs. time plane. This strongly points towards radial migration bringing old [Mg/Fe] and [Fe/H]-rich stars from the inner disk into the APOGEE sample. We learn that when using only chemical evolution modeling we must carefully deconvolve the elemental abundance data from kinematic biases \citep[e.g.][]{Minchev2013}. 
\section{Summary \& Conclusion}
\label{ch:conclusion}
In this paper, we have presented and applied a new modeling tool for galactic chemical evolution studies, \CP . In its basic version, \CP\ is a conventional chemical evolution model: a single-zone "open box". Its new and innovative aspects are centered around efficient Bayesian inference of the model parameters, accounting for a wide range of prior constraints and observations at hand. \CP\ can also be generalized to a multi-zone model, combining "universal" stellar physics with local and diverse ISM histories.
There is a need for such modeling tools, if we are to systematically exploit the ever-growing number of stars with elemental abundance measurements to trace, and ultimately understand, the chemical evolution of galactic systems. 

To illustrate the capabilities of \CP , we used a very limited set of high-quality 
observational constraints: the element abundances of Sun, Arcturus and the local, present-day ISM, traced by  B-stars; we augmented these Galactic constraints, with
broader galaxy population inferences about present-day $L^\star$ galaxies:
their relative incidence of different supernova types and the typical metallicity of their gaseous corona. Using \CP\ we could then show that these data alone already constrain strongly some of fundamental parameters in the Milky Way's chemical evolution. With respect to the question from the sub-title, even the Sun's abundances and its age already precisely determine the IMF and the incidence of SN\,Ia (as can be seen from Figure\,\ref{fig:multiple_obs}).

 In practice, we derived these constraints by placing \CP\  observational predictions into a Bayesian framework, where we marginalize out the model parameters using sampling techniques. We concentrate on three parameters governing the SSP physics (high-mass slope of the IMF, SN\,Ia normalisation and time delay) and four parameters determining the ISM environment (SFE, peak of the SFR, outflow fraction and corona mass), though the method can be easily expanded to constrain more parameters.

We extensively test input and modeling restrictions that inevitably affect 
any such modeling, putting a special focus on \CP 's most important hyperparameter: the yield sets, which translate the chemical evolution prescription of \CP\ into elemental abundance predictions. We implement two different yield sets, consisting of up-to-date yields for the three nucleosynthetic channels: CC-SNe, SNe\,Ia and AGB-winds. Even though the predicted abundance patterns are different for the two yield sets the retrieved chemical evolution parameters are comparable. However, there are persistent discrepancies between the predicted and observed abundances for a number of elements: e.g. we find that K, Ti and V are under-abundant and that Na is over-abundant, irrespective of the chosen yield set and the data sets we try to reproduce. This could hint at a missing nucleosynthetic process \citep{Mernier2016}, or simply reflect shortcomings of the existing yield tables; or it could imply that the net yields are not homogeneously mixed into the ISM. \CP\ can also rank different yield tables by their ability to reproduce elemental abundance data. If the abundances of many stars are used simultaneously, this could be extended to infer yield uncertainty parameters for each element or even to produce metallicity and mass dependent empirical yields.

The basic version of \CP\ can be straightforwardly extended to a multi-zone scheme, where different single-zones share the same SSP physics, but can have separate ISM evolution. We apply this scheme to our set of three stellar abundances and retrieve extremely well-constrained SSP parameters (which we had presumed here to be universal within the Milky Way), together with parameters describing the ISM histories for each star(s): this 
application of \CP\ implies -- unsurprisingly -- that the Sun, B-stars and Arcturus have been experiencing diverse enrichment histories; one might be tempted to associate these with the inner thin disk, outer thin disk and thick disk, respectively. 

With this multi-zone scheme we are, for the first time, able to precisely constrain the IMF high-mass slope from chemical evolution modeling, while properly marginalizing over nuisance parameters, covariances and accounting for the systematic effects of yield tables. The resulting value and achieved precision of $\alpha_\mathrm{IMF}=-2.42^{+0.06}_{-0.05}$ (common uncertainty range of both yield sets) are compatible with the most recent star count analysis in M31 by \cite{Weisz2015}, though we rely on much less data (essentially a few elemental abundances of only three stars), showing the power of the \CP~approach.

Our $\alpha_\mathrm{IMF}$ matches the \citet{Kroupa1993} IMF for the low mass stars (cf. Figure\,\ref{fig:IMF}) and translates to an IMF mass fraction exploding as CC-SNe (i.e. being heavier than 8\,M$_\odot$) of $11.6^{+2.1}_{-1.6}\,\%$ compatible with $14.7^{+2.2}_{-1.0}\,\%$ \citet{Weisz2015} and $12.1\,\%$ \citet{Sa55}. This rules out very steep high-mass slopes \citep{Cz14, Rybizki2015} as well as very shallow ones like the often used \citet{Ch03}. 

The SN\,Ia normalisation from both yield sets ranges from 0.5 to 1.4 events per 1000\,M$_\odot$, which is compatible, but more precise, than the meta-analysis of \citet[tab.~7]{Cote2016b}. The delay time deviates strongly from the prior and turns out very long (3-6\,Gyr), which is the only sensible way within \CP's physical framework to conciliate the $\alpha$-low and $\alpha$-rich abundances. In our particular case, this implication may be driven by the strong $\alpha$-enhancement of Arcturus, despite that fact that it is only $\sim 7$~Gyrs old. But this may also point towards the necessity to include new physics, e.g. more types of SNe \citep{Pakmor2010}.

On the path to fully exploit the chemical imprint of stellar abundance data from large spectroscopic surveys the nucleosynthetic yields will have to be updated and uncertainties added to them, a good start being \citet{Rauscher2016}. Another step will be to take dynamical parameters of the stars into account and get a handle on the age uncertainty, though a statistical age distribution of stellar populations \citep{Just2015a} can be utilized as well. The code is publicly available and the IMF weighted yield output over time (SSP feedback) can be plugged into n-body simulations to use a flexible stellar feedback model. Furthermore new yield tables can be tested quickly and the author is readily available when help with yield table implementation is needed.
\begin{acknowledgements}
We would like to thank the anonymous referee for a thorough reading of the manuscript and many useful comments. The authors would also like to thank David Hogg, Coryn Bailer-Jones, Morgan Fouesneau, Melissa Ness, Fritz R\"opcke, Robert Grand, Greg Stinson, Ivo Seitenzahl and Sam Jones for helpful discussions and valuable feedback on the topic.
JR was partly supported by the DFG Research Centre SFB 881 'The Milky Way System' through project A6. During part of this work JR was a fellow of the International Max Planck Research School for Astronomy and Cosmic Physics at the University of Heidelberg (IMPRS-HD). HWR and JR acknowledge funding from the European Research Council under the
European Union's Seventh Framework Programme (FP 7) ERC Advanced Grant Agreement n. [321035]. HWR acknowledges support of the Miller Institute at UC Berkeley through a visiting professorship during part of this work.
\end{acknowledgements}

\bibliographystyle{aa} 
\bibliography{library} 

\begin{thebibliography}{109}
\expandafter\ifx\csname natexlab\endcsname\relax\def\natexlab#1{#1}\fi

\bibitem[{{Andrews} {et~al.}(2017){Andrews}, {Weinberg}, {Sch{\"o}nrich}, \&
  {Johnson}}]{Andrews2017}
{Andrews}, B.~H., {Weinberg}, D.~H., {Sch{\"o}nrich}, R., \& {Johnson}, J.~A.
  2017, \apj, 835, 224

\bibitem[{{Argast} {et~al.}(2000){Argast}, {Samland}, {Gerhard}, \&
  {Thielemann}}]{Argast2000}
{Argast}, D., {Samland}, M., {Gerhard}, O.~E., \& {Thielemann}, F.-K. 2000,
  \aap, 356, 873

\bibitem[{{Argast} {et~al.}(2002){Argast}, {Samland}, {Thielemann}, \&
  {Gerhard}}]{Argast2001}
{Argast}, D., {Samland}, M., {Thielemann}, F.-K., \& {Gerhard}, O.~E. 2002,
  \aap, 388, 842

\bibitem[{{Asplund} {et~al.}(2009){Asplund}, {Grevesse}, {Sauval}, \&
  {Scott}}]{Asplund2009}
{Asplund}, M., {Grevesse}, N., {Sauval}, A.~J., \& {Scott}, P. 2009, \araa, 47,
  481

\bibitem[{{Baade}(1944)}]{Baade1944}
{Baade}, W. 1944, \apj, 100, 137

\bibitem[{{Basu} \& {Antia}(2004)}]{Basu2004}
{Basu}, S. \& {Antia}, H.~M. 2004, \apjl, 606, L85

\bibitem[{{Battistini} \& {Bensby}(2015)}]{Battistini2015}
{Battistini}, C. \& {Bensby}, T. 2015, \aap, 577, A9

\bibitem[{{Bergemann} {et~al.}(2012){Bergemann}, {Lind}, {Collet}, {Magic}, \&
  {Asplund}}]{Bergemann2012}
{Bergemann}, M., {Lind}, K., {Collet}, R., {Magic}, Z., \& {Asplund}, M. 2012,
  \mnras, 427, 27

\bibitem[{{Bigiel} {et~al.}(2008){Bigiel}, {Leroy}, {Walter}, {Brinks}, {de
  Blok}, {Madore}, \& {Thornley}}]{Bigiel2008}
{Bigiel}, F., {Leroy}, A., {Walter}, F., {et~al.} 2008, \aj, 136, 2846

\bibitem[{{Bovy} {et~al.}(2014){Bovy}, {Nidever}, {Rix}, {Girardi}, {Zasowski},
  {Chojnowski}, {Holtzman}, {Epstein}, {Frinchaboy}, {Hayden}, {Rodrigues},
  {Majewski}, {Johnson}, {Pinsonneault}, {Stello}, {Allende Prieto}, {Andrews},
  {Basu}, {Beers}, {Bizyaev}, {Burton}, {Chaplin}, {Cunha}, {Elsworth},
  {Garc{\'{\i}}a}, {Garc{\'{\i}}a-Her{\'n}andez}, {Garc{\'{\i}}a P{\'e}rez},
  {Hearty}, {Hekker}, {Kallinger}, {Kinemuchi}, {Koesterke},
  {M{\'e}sz{\'a}ros}, {Mosser}, {O'Connell}, {Oravetz}, {Pan}, {Robin},
  {Schiavon}, {Schneider}, {Schultheis}, {Serenelli}, {Shetrone}, {Silva
  Aguirre}, {Simmons}, {Skrutskie}, {Smith}, {Stassun}, {Weinberg}, {Wilson},
  \& {Zamora}}]{Bovy2014b}
{Bovy}, J., {Nidever}, D.~L., {Rix}, H.-W., {et~al.} 2014, \apj, 790, 127

\bibitem[{{Caffau} {et~al.}(2011){Caffau}, {Ludwig}, {Steffen}, {Freytag}, \&
  {Bonifacio}}]{Caffau2011}
{Caffau}, E., {Ludwig}, H.-G., {Steffen}, M., {Freytag}, B., \& {Bonifacio}, P.
  2011, \solphys, 268, 255

\bibitem[{{Chabrier}(2001)}]{Chabrier2001}
{Chabrier}, G. 2001, \apj, 554, 1274

\bibitem[{{Chabrier}(2003)}]{Ch03}
{Chabrier}, G. 2003, \pasp, 115, 763

\bibitem[{{Chiappini} {et~al.}(1997){Chiappini}, {Matteucci}, \&
  {Gratton}}]{Ch97}
{Chiappini}, C., {Matteucci}, F., \& {Gratton}, R. 1997, \apj, 477, 765

\bibitem[{{Chiappini} {et~al.}(2001){Chiappini}, {Matteucci}, \&
  {Romano}}]{Chiappini2001}
{Chiappini}, C., {Matteucci}, F., \& {Romano}, D. 2001, \apj, 554, 1044

\bibitem[{{Chieffi} \& {Limongi}(2004)}]{Chieffi2004}
{Chieffi}, A. \& {Limongi}, M. 2004, \apj, 608, 405

\bibitem[{{C{\^o}t{\'e}} {et~al.}(2017){C{\^o}t{\'e}}, {O'Shea}, {Ritter},
  {Herwig}, \& {Venn}}]{Cote2017}
{C{\^o}t{\'e}}, B., {O'Shea}, B.~W., {Ritter}, C., {Herwig}, F., \& {Venn},
  K.~A. 2017, \apj, 835, 128

\bibitem[{{C{\^o}t{\'e}} {et~al.}(2016{\natexlab{a}}){C{\^o}t{\'e}}, {Ritter},
  {O'Shea}, {Herwig}, {Pignatari}, {Jones}, \& {Fryer}}]{Cote2016b}
{C{\^o}t{\'e}}, B., {Ritter}, C., {O'Shea}, B.~W., {et~al.} 2016{\natexlab{a}},
  \apj, 824, 82

\bibitem[{{C{\^o}t{\'e}} {et~al.}(2016{\natexlab{b}}){C{\^o}t{\'e}}, {West},
  {Heger}, {Ritter}, {O'Shea}, {Herwig}, {Travaglio}, \&
  {Bisterzo}}]{Cote2016f}
{C{\^o}t{\'e}}, B., {West}, C., {Heger}, A., {et~al.} 2016{\natexlab{b}},
  \mnras, 463, 3755

\bibitem[{{Czekaj} {et~al.}(2014){Czekaj}, {Robin}, {Figueras}, {Luri}, \&
  {Haywood}}]{Cz14}
{Czekaj}, M.~A., {Robin}, A.~C., {Figueras}, F., {Luri}, X., \& {Haywood}, M.
  2014, \aap, 564, A102

\bibitem[{{Dziembowski} {et~al.}(1999){Dziembowski}, {Fiorentini}, {Ricci}, \&
  {Sienkiewicz}}]{Dziembowski1998}
{Dziembowski}, W.~A., {Fiorentini}, G., {Ricci}, B., \& {Sienkiewicz}, R. 1999,
  \aap, 343, 990

\bibitem[{{Feuillet} {et~al.}(2016){Feuillet}, {Bovy}, {Holtzman}, {Girardi},
  {MacDonald}, {Majewski}, \& {Nidever}}]{Feuillet2016}
{Feuillet}, D.~K., {Bovy}, J., {Holtzman}, J., {et~al.} 2016, \apj, 817, 40

\bibitem[{{Few} {et~al.}(2012){Few}, {Courty}, {Gibson}, {Kawata}, {Calura}, \&
  {Teyssier}}]{Few2012}
{Few}, C.~G., {Courty}, S., {Gibson}, B.~K., {et~al.} 2012, \mnras, 424, L11

\bibitem[{{Fink} {et~al.}(2014){Fink}, {Kromer}, {Seitenzahl},
  {Ciaraldi-Schoolmann}, {R{\"o}pke}, {Sim}, {Pakmor}, {Ruiter}, \&
  {Hillebrandt}}]{Fink2014}
{Fink}, M., {Kromer}, M., {Seitenzahl}, I.~R., {et~al.} 2014, \mnras, 438, 1762

\bibitem[{Foreman-Mackey(2016)}]{Foreman-Mackey2016}
Foreman-Mackey, D. 2016, The Journal of Open Source Software, 24

\bibitem[{{Foreman-Mackey} {et~al.}(2013){Foreman-Mackey}, {Hogg}, {Lang}, \&
  {Goodman}}]{Foreman2013}
{Foreman-Mackey}, D., {Hogg}, D.~W., {Lang}, D., \& {Goodman}, J. 2013, \pasp,
  125, 306

\bibitem[{{Fox} {et~al.}(2016){Fox}, {Lehner}, {Lockman}, {Wakker}, {Hill},
  {Heitsch}, {Stark}, {Barger}, {Sembach}, \& {Rahman}}]{Fox2016}
{Fox}, A.~J., {Lehner}, N., {Lockman}, F.~J., {et~al.} 2016, \apjl, 816, L11

\bibitem[{{Fran{\c c}ois} {et~al.}(2004){Fran{\c c}ois}, {Matteucci}, {Cayrel},
  {Spite}, {Spite}, \& {Chiappini}}]{Francois2004}
{Fran{\c c}ois}, P., {Matteucci}, F., {Cayrel}, R., {et~al.} 2004, \aap, 421,
  613

\bibitem[{{Goswami} \& {Prantzos}(2000)}]{Goswami2000}
{Goswami}, A. \& {Prantzos}, N. 2000, \aap, 359, 191

\bibitem[{{Grand} {et~al.}(2015){Grand}, {Kawata}, \& {Cropper}}]{Grand2014a}
{Grand}, R.~J.~J., {Kawata}, D., \& {Cropper}, M. 2015, \mnras, 447, 4018

\bibitem[{{Henry} {et~al.}(2010){Henry}, {Cowan}, \& {Sobeck}}]{Henry2010}
{Henry}, R.~B.~C., {Cowan}, J.~J., \& {Sobeck}, J. 2010, \apj, 709, 715

\bibitem[{{Henry} {et~al.}(2000){Henry}, {Edmunds}, \&
  {K{\"o}ppen}}]{Henry2000}
{Henry}, R.~B.~C., {Edmunds}, M.~G., \& {K{\"o}ppen}, J. 2000, \apj, 541, 660

\bibitem[{{Iben}(1965)}]{Iben1965}
{Iben}, Jr., I. 1965, \apj, 142, 1447

\bibitem[{{Iwamoto} {et~al.}(1999){Iwamoto}, {Brachwitz}, {Nomoto},
  {Kishimoto}, {Umeda}, {Hix}, \& {Thielemann}}]{Iwamoto1999}
{Iwamoto}, K., {Brachwitz}, F., {Nomoto}, K., {et~al.} 1999, \apjs, 125, 439

\bibitem[{{Jim{\'e}nez} {et~al.}(2015){Jim{\'e}nez}, {Tissera}, \&
  {Matteucci}}]{Jimenez2015}
{Jim{\'e}nez}, N., {Tissera}, P.~B., \& {Matteucci}, F. 2015, \apj, 810, 137

\bibitem[{{Jofre} {et~al.}(2016){Jofre}, {Heiter}, {Worley}, {Blanco-Cuaresma},
  {Soubiran}, {Masseron}, {Hawkins}, {Adibekyan}, {Buder}, {Casamiquela},
  {Gilmore}, {Hourihane}, \& {Tabernero}}]{Jofre2016}
{Jofre}, P., {Heiter}, U., {Worley}, C.~C., {et~al.} 2016
  [\eprint[arXiv]{1612.05013}]

\bibitem[{{J{\"o}nsson} {et~al.}(2014){J{\"o}nsson}, {Ryde}, {Harper},
  {Richter}, \& {Hinkle}}]{Jonsson2014}
{J{\"o}nsson}, H., {Ryde}, N., {Harper}, G.~M., {Richter}, M.~J., \& {Hinkle},
  K.~H. 2014, \apjl, 789, L41

\bibitem[{{Just} \& {Rybizki}(2016)}]{Just2015a}
{Just}, A. \& {Rybizki}, J. 2016, Astronomische Nachrichten, 337, 880

\bibitem[{{Karakas}(2010)}]{Karakas2010}
{Karakas}, A.~I. 2010, \mnras, 403, 1413

\bibitem[{{Karakas} \& {Lugaro}(2016)}]{Karakas2016}
{Karakas}, A.~I. \& {Lugaro}, M. 2016, \apj, 825, 26

\bibitem[{{Kobayashi} {et~al.}(2006){Kobayashi}, {Umeda}, {Nomoto}, {Tominaga},
  \& {Ohkubo}}]{Kobayashi2006}
{Kobayashi}, C., {Umeda}, H., {Nomoto}, K., {Tominaga}, N., \& {Ohkubo}, T.
  2006, \apj, 653, 1145

\bibitem[{{Kroupa} {et~al.}(1993){Kroupa}, {Tout}, \& {Gilmore}}]{Kroupa1993}
{Kroupa}, P., {Tout}, C.~A., \& {Gilmore}, G. 1993, \mnras, 262, 545

\bibitem[{{Kubryk} {et~al.}(2015){Kubryk}, {Prantzos}, \&
  {Athanassoula}}]{Kubryk2015}
{Kubryk}, M., {Prantzos}, N., \& {Athanassoula}, E. 2015, \aap, 580, A126

\bibitem[{{Lawler} {et~al.}(1989){Lawler}, {Brownlee}, {Temple}, \&
  {Wheelock}}]{Lawler1989}
{Lawler}, M.~E., {Brownlee}, D.~E., {Temple}, S., \& {Wheelock}, M.~M. 1989,
  \icarus, 80, 225

\bibitem[{{Lodders} {et~al.}(2009){Lodders}, {Palme}, \& {Gail}}]{Lodders2009}
{Lodders}, K., {Palme}, H., \& {Gail}, H.-P. 2009, Landolt B{\"o}rnstein

\bibitem[{{Maeder}(1992)}]{Maeder1992}
{Maeder}, A. 1992, \aap, 264, 105

\bibitem[{{Majewski} {et~al.}(2016){Majewski}, {APOGEE Team}, \& {APOGEE-2
  Team}}]{Majewski2016}
{Majewski}, S.~R., {APOGEE Team}, \& {APOGEE-2 Team}. 2016, Astronomische
  Nachrichten, 337, 863

\bibitem[{{Mannucci} {et~al.}(2005){Mannucci}, {Della Valle}, {Panagia},
  {Cappellaro}, {Cresci}, {Maiolino}, {Petrosian}, \& {Turatto}}]{Mannucci2005}
{Mannucci}, F., {Della Valle}, M., {Panagia}, N., {et~al.} 2005, \aap, 433, 807

\bibitem[{{Maoz} \& {Mannucci}(2012)}]{Maoz2012a}
{Maoz}, D. \& {Mannucci}, F. 2012, \pasa, 29, 447

\bibitem[{{Maoz} {et~al.}(2012){Maoz}, {Mannucci}, \& {Brandt}}]{Maoz2012}
{Maoz}, D., {Mannucci}, F., \& {Brandt}, T.~D. 2012, \mnras, 426, 3282

\bibitem[{{Maoz} {et~al.}(2010){Maoz}, {Sharon}, \& {Gal-Yam}}]{Maoz2010}
{Maoz}, D., {Sharon}, K., \& {Gal-Yam}, A. 2010, \apj, 722, 1879

\bibitem[{{Martig} {et~al.}(2016){Martig}, {Fouesneau}, {Rix}, {Ness},
  {M{\'e}sz{\'a}ros}, {Garc{\'{\i}}a-Hern{\'a}ndez}, {Pinsonneault},
  {Serenelli}, {Silva Aguirre}, \& {Zamora}}]{Martig2016}
{Martig}, M., {Fouesneau}, M., {Rix}, H.-W., {et~al.} 2016, \mnras, 456, 3655

\bibitem[{{Matteucci}(2003)}]{Matteucci2001}
{Matteucci}, F. 2003, Astrophysics and Space Science Library, Vol. 253, {The
  Chemical Evolution of the Galaxy} (Kluwer Academic Publishers, Dordrecht)

\bibitem[{{Matteucci}(2012)}]{Matteucci2012}
{Matteucci}, F. 2012, {Chemical Evolution of Galaxies}, Astronomy and
  Astrophysics Library (Springer-Verlag Berlin Heidelberg)

\bibitem[{{Matteucci} \& {Recchi}(2001)}]{Matteucci2001a}
{Matteucci}, F. \& {Recchi}, S. 2001, \apj, 558, 351

\bibitem[{{McDonough}(1995)}]{McDonough1995}
{McDonough}, W.~F. 1995, in Lunar and Planetary Science Conference, Vol.~26

\bibitem[{{Mernier} {et~al.}(2016){Mernier}, {de Plaa}, {Pinto}, {Kaastra},
  {Kosec}, {Zhang}, {Mao}, \& {Werner}}]{Mernier2016}
{Mernier}, F., {de Plaa}, J., {Pinto}, C., {et~al.} 2016, \aap, 592, A157

\bibitem[{{Mikolaitis} {et~al.}(2016){Mikolaitis}, {de Laverny},
  {Recio-Blanco}, {Hill}, {Worley}, \& {de Pascale}}]{Mikolaitis2016}
{Mikolaitis}, {\v S}., {de Laverny}, P., {Recio-Blanco}, A., {et~al.} 2016
  [\eprint[arXiv]{1612.07622}]

\bibitem[{{Minchev} {et~al.}(2013){Minchev}, {Chiappini}, \&
  {Martig}}]{Minchev2013}
{Minchev}, I., {Chiappini}, C., \& {Martig}, M. 2013, \aap, 558, A9

\bibitem[{{M{\"u}ller}(2016)}]{Muller2016}
{M{\"u}ller}, B. 2016, \pasa, 33, e048

\bibitem[{{Ness} {et~al.}(2016){Ness}, {Hogg}, {Rix}, {Martig}, {Pinsonneault},
  \& {Ho}}]{Ness2016}
{Ness}, M., {Hogg}, D.~W., {Rix}, H.-W., {et~al.} 2016, \apj, 823, 114

\bibitem[{{Nieva} \& {Przybilla}(2012)}]{Nieva2012}
{Nieva}, M.-F. \& {Przybilla}, N. 2012, \aap, 539, A143

\bibitem[{{Nomoto} {et~al.}(2013){Nomoto}, {Kobayashi}, \&
  {Tominaga}}]{Nomoto2013}
{Nomoto}, K., {Kobayashi}, C., \& {Tominaga}, N. 2013, \araa, 51, 457

\bibitem[{{Pakmor} {et~al.}(2010){Pakmor}, {Kromer}, {R{\"o}pke}, {Sim},
  {Ruiter}, \& {Hillebrandt}}]{Pakmor2010}
{Pakmor}, R., {Kromer}, M., {R{\"o}pke}, F.~K., {et~al.} 2010, \nat, 463, 61

\bibitem[{{Pignatari} {et~al.}(2016){Pignatari}, {Herwig}, {Hirschi},
  {Bennett}, {Rockefeller}, {Fryer}, {Timmes}, {Ritter}, {Heger}, {Jones},
  {Battino}, {Dotter}, {Trappitsch}, {Diehl}, {Frischknecht}, {Hungerford},
  {Magkotsios}, {Travaglio}, \& {Young}}]{Pignatari2016}
{Pignatari}, M., {Herwig}, F., {Hirschi}, R., {et~al.} 2016, \apjs, 225, 24

\bibitem[{{Pilachowski} \& {Pace}(2015)}]{Pilachowski2015}
{Pilachowski}, C.~A. \& {Pace}, C. 2015, \aj, 150, 66

\bibitem[{{Portinari} {et~al.}(1998){Portinari}, {Chiosi}, \&
  {Bressan}}]{Portinari1998}
{Portinari}, L., {Chiosi}, C., \& {Bressan}, A. 1998, \aap, 334, 505

\bibitem[{{Ram{\'{\i}}rez} \& {Allende Prieto}(2011)}]{Ramirez2011}
{Ram{\'{\i}}rez}, I. \& {Allende Prieto}, C. 2011, \apj, 743, 135

\bibitem[{{Rauscher} {et~al.}(2016){Rauscher}, {Nishimura}, {Hirschi},
  {Cescutti}, {Murphy}, \& {Heger}}]{Rauscher2016}
{Rauscher}, T., {Nishimura}, N., {Hirschi}, R., {et~al.} 2016, \mnras, 463,
  4153

\bibitem[{{Recio-Blanco} {et~al.}(2012){Recio-Blanco}, {de Laverny}, {Worley},
  {Santos}, {Melo}, \& {Israelian}}]{Recio-Blanco2012}
{Recio-Blanco}, A., {de Laverny}, P., {Worley}, C., {et~al.} 2012, \aap, 538,
  A117

\bibitem[{{Reeves}(1970)}]{Reeves1970}
{Reeves}, H. 1970, \nat, 226, 727

\bibitem[{{Romano} {et~al.}(2010){Romano}, {Karakas}, {Tosi}, \&
  {Matteucci}}]{Romano2010}
{Romano}, D., {Karakas}, A.~I., {Tosi}, M., \& {Matteucci}, F. 2010, \aap, 522,
  A32

\bibitem[{{Romano} {et~al.}(1999){Romano}, {Matteucci}, {Molaro}, \&
  {Bonifacio}}]{Romano1999}
{Romano}, D., {Matteucci}, F., {Molaro}, P., \& {Bonifacio}, P. 1999, \aap,
  352, 117

\bibitem[{{Ruiter} {et~al.}(2009){Ruiter}, {Belczynski}, \&
  {Fryer}}]{Ruiter2009}
{Ruiter}, A.~J., {Belczynski}, K., \& {Fryer}, C. 2009, \apj, 699, 2026

\bibitem[{{Rybizki} \& {Just}(2015)}]{Rybizki2015}
{Rybizki}, J. \& {Just}, A. 2015, \mnras, 447, 3880

\bibitem[{{Salpeter}(1955)}]{Sa55}
{Salpeter}, E.~E. 1955, \apj, 121, 161

\bibitem[{{Schmidt}(1959)}]{Schmidt1959}
{Schmidt}, M. 1959, \apj, 129, 243

\bibitem[{{Sch{\"o}nrich} \& {Binney}(2009)}]{Schonrich2009}
{Sch{\"o}nrich}, R. \& {Binney}, J. 2009, \mnras, 396, 203

\bibitem[{{SDSS Collaboration} {et~al.}(2016){SDSS Collaboration}, {Albareti},
  {Allende Prieto}, {Almeida}, {Anders}, {Anderson}, {Andrews},
  {Aragon-Salamanca}, {Argudo-Fernandez}, {Armengaud}, \&
  et~al.}]{SDSSCollaboration2016}
{SDSS Collaboration}, {Albareti}, F.~D., {Allende Prieto}, C., {et~al.} 2016
  [\eprint[arXiv]{1608.02013}]

\bibitem[{{Seitenzahl} {et~al.}(2013){Seitenzahl}, {Ciaraldi-Schoolmann},
  {R{\"o}pke}, {Fink}, {Hillebrandt}, {Kromer}, {Pakmor}, {Ruiter}, {Sim}, \&
  {Taubenberger}}]{Seitenzahl2013}
{Seitenzahl}, I.~R., {Ciaraldi-Schoolmann}, F., {R{\"o}pke}, F.~K., {et~al.}
  2013, \mnras, 429, 1156

\bibitem[{{Serenelli}(2016)}]{Serenelli2016}
{Serenelli}, A. 2016, European Physical Journal A, 52, 78

\bibitem[{{Sim} {et~al.}(2013){Sim}, {Seitenzahl}, {Kromer},
  {Ciaraldi-Schoolmann}, {R{\"o}pke}, {Fink}, {Hillebrandt}, {Pakmor},
  {Ruiter}, \& {Taubenberger}}]{Sim2013}
{Sim}, S.~A., {Seitenzahl}, I.~R., {Kromer}, M., {et~al.} 2013, \mnras, 436,
  333

\bibitem[{{Smith}(1963)}]{Smith1963}
{Smith}, G.~P. 1963, \bain, 17, 203

\bibitem[{{Snaith} {et~al.}(2014){Snaith}, {Haywood}, {Di Matteo}, {Lehnert},
  {Combes}, {Katz}, \& {G{\'o}mez}}]{Snaith2014a}
{Snaith}, O.~N., {Haywood}, M., {Di Matteo}, P., {et~al.} 2014, \apjl, 781, L31

\bibitem[{{Spitoni} {et~al.}(2017){Spitoni}, {Vincenzo}, \&
  {Matteucci}}]{Spitoni2017}
{Spitoni}, E., {Vincenzo}, F., \& {Matteucci}, F. 2017, \aap, 599, A6

\bibitem[{{Steffen} {et~al.}(2015){Steffen}, {Prakapavi{\v c}ius}, {Caffau},
  {Ludwig}, {Bonifacio}, {Cayrel}, {Ku{\v c}inskas}, \&
  {Livingston}}]{Steffen2015}
{Steffen}, M., {Prakapavi{\v c}ius}, D., {Caffau}, E., {et~al.} 2015, \aap,
  583, A57

\bibitem[{{Stern} {et~al.}(2016){Stern}, {Hennawi}, {Prochaska}, \&
  {Werk}}]{Stern2016}
{Stern}, J., {Hennawi}, J.~F., {Prochaska}, J.~X., \& {Werk}, J.~K. 2016, \apj,
  830, 87

\bibitem[{{Stinson} {et~al.}(2006){Stinson}, {Seth}, {Katz}, {Wadsley},
  {Governato}, \& {Quinn}}]{Stinson2006}
{Stinson}, G., {Seth}, A., {Katz}, N., {et~al.} 2006, \mnras, 373, 1074

\bibitem[{{Sukhbold} {et~al.}(2016){Sukhbold}, {Ertl}, {Woosley}, {Brown}, \&
  {Janka}}]{Sukhbold2015}
{Sukhbold}, T., {Ertl}, T., {Woosley}, S.~E., {Brown}, J.~M., \& {Janka}, H.-T.
  2016, \apj, 821, 38

\bibitem[{{Thielemann} {et~al.}(2003){Thielemann}, {Argast}, {Brachwitz},
  {Hix}, {H{\"o}flich}, {Liebend{\"o}rfer}, {Martinez-Pinedo}, {Mezzacappa},
  {Panov}, \& {Rauscher}}]{Thielemann2003}
{Thielemann}, F.-K., {Argast}, D., {Brachwitz}, F., {et~al.} 2003, \nphysa,
  718, 139

\bibitem[{{Timmes} {et~al.}(1995){Timmes}, {Woosley}, \& {Weaver}}]{Timmes1995}
{Timmes}, F.~X., {Woosley}, S.~E., \& {Weaver}, T.~A. 1995, \apjs, 98, 617

\bibitem[{{Tinsley}(1979)}]{Tinsley1979}
{Tinsley}, B.~M. 1979, \apj, 229, 1046

\bibitem[{{Tinsley}(1980)}]{Tinsley1980}
{Tinsley}, B.~M. 1980, \fcp, 5, 287

\bibitem[{{Turcotte} \& {Wimmer-Schweingruber}(2002)}]{Turcotte2002}
{Turcotte}, S. \& {Wimmer-Schweingruber}, R.~F. 2002, Journal of Geophysical
  Research (Space Physics), 107, 1442

\bibitem[{{Valls-Gabaud}(2014)}]{Valls-Gabaud2014}
{Valls-Gabaud}, D. 2014, in EAS Publications Series, Vol.~65, 225--265

\bibitem[{{van den Hoek} \& {Groenewegen}(1997)}]{VandenHoek1997}
{van den Hoek}, L.~B. \& {Groenewegen}, M.~A.~T. 1997, \aaps, 123

\bibitem[{{van der Walt} {et~al.}(2011){van der Walt}, {Colbert}, \&
  {Varoquaux}}]{VanderWalt2011}
{van der Walt}, S., {Colbert}, C., \& {Varoquaux}, G. 2011, Computing in
  Science \& Engineering, 13, 22

\bibitem[{{van Dokkum} {et~al.}(2013){van Dokkum}, {Leja}, {Nelson}, {Patel},
  {Skelton}, {Momcheva}, {Brammer}, {Whitaker}, {Lundgren}, {Fumagalli},
  {Conroy}, {F{\"o}rster Schreiber}, {Franx}, {Kriek}, {Labb{\'e}},
  {Marchesini}, {Rix}, {van der Wel}, \& {Wuyts}}]{VanDokkum2013b}
{van Dokkum}, P.~G., {Leja}, J., {Nelson}, E.~J., {et~al.} 2013, \apjl, 771,
  L35

\bibitem[{{Venn} {et~al.}(2012){Venn}, {Shetrone}, {Irwin}, {Hill}, {Jablonka},
  {Tolstoy}, {Lemasle}, {Divell}, {Starkenburg}, {Letarte}, {Baldner},
  {Battaglia}, {Helmi}, {Kaufer}, \& {Primas}}]{Venn2012}
{Venn}, K.~A., {Shetrone}, M.~D., {Irwin}, M.~J., {et~al.} 2012, \apj, 751, 102

\bibitem[{{Ventura} {et~al.}(2012){Ventura}, {D'Antona}, {Di Criscienzo},
  {Carini}, {D'Ercole}, \& {vesperini}}]{Ventura2012}
{Ventura}, P., {D'Antona}, F., {Di Criscienzo}, M., {et~al.} 2012, \apjl, 761,
  L30

\bibitem[{{Ventura} {et~al.}(2013){Ventura}, {Di Criscienzo}, {Carini}, \&
  {D'Antona}}]{Ventura2013}
{Ventura}, P., {Di Criscienzo}, M., {Carini}, R., \& {D'Antona}, F. 2013,
  \mnras, 431, 3642

\bibitem[{{Vincenzo} {et~al.}(2017){Vincenzo}, {Matteucci}, \&
  {Spitoni}}]{Vincenzo2016b}
{Vincenzo}, F., {Matteucci}, F., \& {Spitoni}, E. 2017, \mnras, 466, 2939

\bibitem[{{Wallerstein}(1962)}]{Wallerstein1962}
{Wallerstein}, G. 1962, \apjs, 6, 407

\bibitem[{{Weinberg} {et~al.}(2016){Weinberg}, {Andrews}, \&
  {Freudenburg}}]{Weinberg2016}
{Weinberg}, D.~H., {Andrews}, B.~H., \& {Freudenburg}, J. 2016
  [\eprint[arXiv]{1604.07435}]

\bibitem[{{Weisz} {et~al.}(2015){Weisz}, {Johnson}, {Foreman-Mackey},
  {Dolphin}, {Beerman}, {Williams}, {Dalcanton}, {Rix}, {Hogg}, {Fouesneau},
  {Johnson}, {Bell}, {Boyer}, {Gouliermis}, {Guhathakurta}, {Kalirai}, {Lewis},
  {Seth}, \& {Skillman}}]{Weisz2015}
{Weisz}, D.~R., {Johnson}, L.~C., {Foreman-Mackey}, D., {et~al.} 2015, \apj,
  806, 198

\bibitem[{{Werk} {et~al.}(2014){Werk}, {Prochaska}, {Tumlinson}, {Peeples},
  {Tripp}, {Fox}, {Lehner}, {Thom}, {O'Meara}, {Ford}, {Bordoloi}, {Katz},
  {Tejos}, {Oppenheimer}, {Dav{\'e}}, \& {Weinberg}}]{Werk2014}
{Werk}, J.~K., {Prochaska}, J.~X., {Tumlinson}, J., {et~al.} 2014, \apj, 792, 8

\bibitem[{{Wielen} {et~al.}(1996){Wielen}, {Fuchs}, \& {Dettbarn}}]{Wielen1996}
{Wielen}, R., {Fuchs}, B., \& {Dettbarn}, C. 1996, \aap, 314, 438

\bibitem[{{Woosley} \& {Weaver}(1994)}]{Woosley1994}
{Woosley}, S.~E. \& {Weaver}, T.~A. 1994, \apj, 423, 371

\bibitem[{{Woosley} \& {Weaver}(1995)}]{Woosley1995}
{Woosley}, S.~E. \& {Weaver}, T.~A. 1995, \apjs, 101, 181

\end{thebibliography}

\end{document}